\documentclass[11pt]{article}
\usepackage{amsmath,amssymb,color,epsfig,cite}
\usepackage{graphicx}
\usepackage{subfigure}
\usepackage{setspace}

\usepackage{amsfonts}

\newcommand{\hoch}[1]{$\, ^{#1}$}

\makeatletter
\@addtoreset{equation}{section}
\makeatother

\def\crampest{\medmuskip = 1mu plus 1mu minus 1mu}
\def\uncramp{\medmuskip = 4mu plus 2mu minus 4mu}

\newcommand{\be}{\begin{equation}}
\newcommand{\ee}{\end{equation}}
\newcommand{\bea}{\setlength\arraycolsep{2pt} \begin{eqnarray}}
\newcommand{\eea}{\end{eqnarray}}
\newcommand{\nn}{\nonumber}

\def\ft#1#2{{\textstyle{\frac{\scriptstyle #1}{\scriptstyle #2} } }}
\def\fft#1#2{{\frac{#1}{#2}}}

\def\0{{\sst{(0)}}}
\def\1{{\sst{(1)}}}
\def\2{{\sst{(2)}}}
\def\3{{\sst{(3)}}}
\def\4{{\sst{(4)}}}
\def\5{{\sst{(5)}}}
\def\6{{\sst{(6)}}}
\def\7{{\sst{(7)}}}
\def\8{{\sst{(8)}}}
\def\sst#1{{\scriptscriptstyle #1}}

\def\arcsinh{{\rm arcsinh}}

\begin{document}

\begin{flushright}
\hfill{ MI-TH-187}

\end{flushright}

\begin{center}
{\large {\bf Mass of Dyonic Black Holes and Entropy Super-Additivity}}

\vspace{15pt}
{\large
Wei-Jian Geng\hoch{1\dagger}, Blake Giant\hoch{2\ddagger}, H. L\"u\hoch{3*} and
C.N. Pope\hoch{4,5\sharp}}

\vspace{15pt}

\hoch{1}{\it Department of Physics, Beijing Normal University, Beijing 100875, China}
\vspace{10pt}

\hoch{2}{\it Department of Physics, Lehigh University, Bethlehem, PA 18018,
  USA}

\vspace{10pt}

\hoch{3}{\it Center for Joint Quantum Studies, School of Science,\\
 Tianjin University, Tianjin 300350, China}

\vspace{10pt}

\hoch{4}{\it George P. \& Cynthia Woods Mitchell  Institute
for Fundamental Physics and Astronomy,\\
Texas A\&M University, College Station, TX 77843, USA}

\vspace{10pt}

\hoch{5}{\it DAMTP, Centre for Mathematical Sciences,
 Cambridge University,\\  Wilberforce Road, Cambridge CB3 OWA, UK}

\vspace{30pt}

\underline{ABSTRACT}
\end{center}

   We study extremal static dyonic black holes in four-dimensional
Einstein-Maxwell-Dilaton theory, for general values of the
constant $a$ in the exponential coupling $e^{a\phi}$ of the dilaton to
the Maxwell kinetic term.  Explicit solutions are known only for $a=0$,
$a=1$ and $a=\sqrt3$, and for general $a$ when the electric and magnetic
charges $Q$ and $P$ are equal.  We obtain solutions as power series
expansions around $Q=P$, in terms of a small parameter $\epsilon=
a^{-1}\, \log(Q/P)$.  Using these, and also solutions constructed
numerically, we test a relation between the mass and the
charges that had been conjectured long ago by Rasheed.  We find that although
the conjecture is not exactly correct it is in fact quite accurate for
a wide range of the black hole parameters. We investigate some improved
conjectures for the mass relation.  We also study the circumstances
under which entropy super-additivity, which is related to Hawking's
area theorem, is violated.  This extends beyond previous examples
exhibited in the literature for the particular case of $a=\sqrt3$ dyonic
black holes.

\vfill
{\footnotesize \noindent\hoch{\dagger}gengwj@mail.bnu.edu.cn\  \ \hoch{\ddagger}bkg218@lehigh.edu\ \
\hoch{*}mrhonglu@gmail.com\ \ \hoch{\sharp}pope@physics.tamu.edu}

\thispagestyle{empty}

\pagebreak

\tableofcontents
\addtocontents{toc}{\protect\setcounter{tocdepth}{2}}


\newpage

\section{Introduction}

    The Einstein-Maxwell-Dilaton (EMD) theory in four dimensions, and its
black hole solutions, have been investigated extensively in the
last few decades.  The theory
 is a generalization of Einstein-Maxwell theory in
which a real dilatonic scalar field $\phi$ is included,
coupling not only to gravity but also, via a non-minimal term,
to the Maxwell field. The Lagrangian is given by
\be
{\cal L}=\sqrt{-g} \Big( R - \ft12 (\partial\phi)^2 -
\ft14 e^{a\phi} F^2\Big)\,,\qquad F = dA \,. \label{EMD}
\ee
For certain specific values of the dilaton coupling constant $a$, the
EMD theory corresponds to particular truncations of four-dimensional
maximal ${\cal N}=8$ supergravity, and for this reason much attention has been
focused on these cases because of their relevance in string theory.
In fact, these special cases all fall within the consistent truncation
of the ${\cal N}=8$ theory to ${\cal N}=2$ STU supergravity \cite{Duff:1995sm}, which
comprises pure ${\cal N}=2$ supergravity coupled to three vector multiplets.
The further truncations to special cases of the EMD theory arise by setting
some of the STU supergravity fields, or combinations of fields, to zero.  The
various truncations lead to an EMD theory with $a$ taking one of the
values $a=0$, $\ft1{\sqrt3}$, 1 or $\sqrt3$ \cite{Duff:1994jr,Gibbons:1994vm,Lu:1995yn}. 
The case of $a=0$ can also
be reduced to pure Einstein-Maxwell theory, since the dilaton can be
consistently truncated in this case.  When $a=\sqrt3$ the EMD theory also
corresponds
to the Kaluza-Klein reduction of five-dimensional pure gravity.  Because of
the enhanced Cremer-Julia like symmetries \cite{Cremmer:1978ds,cjgroup,
Cremmer:1997ct,Cremmer:1998px}
of the dimensional reduction of the supergravity theories, in these four 
special cases among the general EMD theories, one can use solution-generating 
techniques to construct charged black hole solutions from neutral ones.

   Static non-extremal black holes in the general four-dimensional EMD
theory were constructed in \cite{gibmae,gibwil}.  Explicit solutions are
obtainable for all values of the dilaton coupling in the case where the
black hole carries purely electric or purely magnetic charge.  Dyonic
black holes, carrying both electric and magnetic charge, can be constructed
explicitly in the three cases $a=0$, $a=1$ and $a=\sqrt3$.  When $a=0$
this is rather trivial, since there then exists a duality symmetry of the
equations of motion, allowing a purely electric or purely magnetic
Reissner-Nordstr\"om (RN) black hole to be transformed into a dyonic one.  Rotating
black holes have also been studied in the EMD theory.  In particular,
the dyonic rotating solutions in the $a=\sqrt3$ EMD theory were constructed
by Rasheed in \cite{rasheed} (see also \cite{larsen}).

   In this paper we shall focus our attention on static black hole
solutions in the EMD theories; in particular on the dyonically-charged
black holes.  As mentioned above, these can be constructed explicitly
when $a=0$, 1 or $\sqrt3$, but no explicit solutions
are known for other values of the dilaton coupling, except when the electric
and magnetic charges are equal, in which case the dilaton is constant and
the solution reduces to a dyonic Reissner-Nordstr\"om black hole.
We shall therefore
use various approximation and numerical techniques in order to investigate
the nature of the dyonic black hole solutions for general values of $a$.
We shall, furthermore, restrict our attention to the case of {\it extremal}
dyonic black holes.

  In the extremal limit, the mass of a black hole is no longer specifiable
independently, but rather, it is a definite function of the electric and
magnetic charges.  For the three special cases mentioned above, where
explicit solutions are known, the mass $M$ is given in terms of the
electric charge $Q$ and magnetic charge $P$ by the relations
\bea
a = 0 \,, &&\quad M = 2 \sqrt{Q^2 + P^2}\,, \cr
a = 1 \,, &&\quad M = \sqrt{2}(Q + P)\,, \cr
a = \sqrt{3} \,, &&\quad M = (Q^\fft23 + P^\fft23)^\fft32 \,.\label{relation1}
\eea
Based on these expressions, and also the known mass-charge relations of
extremal single-charge or equal-charge solutions, Rasheed
\cite{rasheed} conjectured a possible formula
for the mass as a function of $Q$ and $P$ for dyonic extremal black holes
for general values of the dilaton coupling constant $a$, namely
\be
M=\fft2{\sqrt{a^2 + 1}} (Q^b + P^b)^{\fft{1}{b}}\,,\qquad
b=\fft{2\log2}{\log(2a^2 + 2)}\,.\label{rasheedconj}
\ee
The approximate solutions that we are able to construct for general values
of the dilaton coupling allow us to test the validity of this conjecture, and
in fact, we are able to establish that it does not hold. However, we are able to
obtain an approximate relation, valid in the regime when $Q$ and $P$ are
close to one another, but we have not found a
simple closed-form relation for $M=M(Q,P)$ that is valid for general
values of the dilaton coupling $a$.

  The other purpose of this paper is to study a thermodynamic
property of black holes known as entropy super-additivity or
sub-additivity.  Gravitating systems, and black holes especially,
have thermodynamic properties that are rather different from those of more
conventional laboratory systems.  This is related to the the long-range
nature of the gravitational field, and the fact that it cannot be screened.
In particular, the entropy does not scale homogeneously with the mass of
the black hole.  Unlike in more conventional thermodynamic systems, the
entropy of a black hole is not a concave function of the other extensive
variables.  A property that {\it does} hold, however, in many black hole
solutions, is that the entropy is a super-additive function. That is,
if one considers black hole solutions with mass $M$, angular momentum $J$,
charges $Q^i$, and entropy $S=S(M,J,Q^i)$, then for two such
sets of parameters $(M_1,J_1,Q^i_1)$ and $(M_2,J_2,Q^i_2)$, entropy
super-additivity asserts that
\be
S(M_1+M_2,J_1+J_2,Q^i_1+Q^i_2) \ge
    S(M_1,J_1,Q^i_1) + S(M_2,J_2,Q^i_2)\,,\label{Ssuperadd}
\ee
(with all angular momenta and charges taken, without loss of generality, to
be non-negative).  This property was observed for Kerr-Newman black holes
in \cite{tranland}, and was recently explored for a variety of black holes
in \cite{cvgiluponeg}.  Situations where entropy super-additivity is not
satisfied seem to be rather rare, and just one example was
exhibited in \cite{cvgiluponeg}, namely the dyonic black hole in $a=\sqrt3$
EMD theory.  Since entropy super-additivity holds for the other
explicitly known examples of dyonic EMD black holes (i.e.~for $a=0$ and $a=1$),
it is therefore of considerable interest to investigate this question for
the case of general dilaton coupling $a$.

The paper is organized as follows:  In section 2 we obtain the
equations of motion for the various functions in the ansatz for static
dyonic solutions, including the equation relating the metric functions in
the case of extremal solutions.  In section 3 we discuss the known
explicit static solutions, including purely electric and purely magnetic black
holes for all values of the dilaton coupling $a$, and the dyonic
black hole solutions for the special cases $a=0$, $a=1$ and $a=\sqrt3$.
In section 4 we present our construction of extremal dyonic solutions for
general dilaton coupling $a$, as power series expansions in the
parameter $\epsilon=a^{-1}\, \log(Q/P)$, by perturbing around the
exactly-solvable case where $Q=P$.  We then use these solutions to
investigate candidate formulae giving the mass in terms of the electric
and magnetic charges of the extremal black holes.  We also test the
mass formulae in the regime far away from $Q\sim P$, by constructing
numerical solutions for the dyonic black holes.  In section 5 we examine
the conditions under which entropy super-additivity holds, showing that
counter-examples can arise when $a$ is sufficiently large.  Finally, after
conclusions in section 6, we give some further details of our series
expansion results in appendix A, and our numerical procedures in
appendix B.

\section{The theory and equations of motion}

In this section, we study the EMD theory (\ref{EMD}),
focusing on constructing spherically symmetric black hole solutions.
It is sometimes convenient
to express the dilaton coupling constant $a$ as \cite{Lu:1995yn}
\be
a^2 = \fft{4}{N} - 1\,.
\ee
When $N=1$, 2, 3 or 4, corresponding to $a=\sqrt3$, 1, $\ft{1}{\sqrt3}$ or 0,
the theory can be consistently embedded in a supergravity theory,
with $N$ denoting the number of basic stringy building blocks in string
or M-theory \cite{Lu:1995yn}. The equations of motion are
\bea
\Box\phi &=& \ft14 a e^{a\phi} F^2\,,\qquad \nabla_\mu e^{a\phi} F^{\mu\nu} = 0\,,\cr
E_{\mu\nu} &\equiv& R_{\mu\nu} - \ft12 g_{\mu\nu} R = \ft12 \left(\partial_\mu \phi \partial_\nu\phi
-\ft12 g_{\mu\nu} (\partial\phi)^2\right) +\ft12 e^{a\phi} \left(F_{\mu\nu}^2 - \ft14 g_{\mu\nu} F^2\right)\,.
\eea

We shall consider static, spherically-symmetric solutions.
The most general ansatz with both electric and magnetic charges is
\bea
ds^2 &=& - h(r) dt^2 + \fft{dr^2}{f(r)} + r^2 d\Omega^2\,,\qquad
\phi=\phi(r)\,,\cr
A &=& A_0(r) dt + p \cos\theta\, d\phi\,, \label{ansatz}
\eea
where $p$ is a constant, associated with the magnetic charge. The Maxwell equation implies that
\be
A_0'=\fft{q}{r^2}\,e^{-a\phi}\,\sqrt{\fft{h}{f}}\,,
\ee
where the integration constant $q$ is proportional to the electric charge.
The electric and magnetic charges are given by
\be
Q=\fft{1}{16\pi} \int_{r\rightarrow \infty} e^{a\phi} {* F} =\ft14 q\,,\qquad
P=\fft{1}{16\pi} \int F = \ft14 p\,.
\ee
The scalar equation of motion is then given by
\be
\phi'' + \ft12 \Big(\fft{h'}{h} + \fft{f'}{f} + \fft{4}{r}\Big) \phi'
- \fft{a}{2r^4f} (p^2 e^{a\phi} - q^2 e^{-a\phi})=0\,.
\ee
The Einstein equations of motion can be given in terms of the combinations
\bea
E^{t}_t - E^{r}_r &=& \fft{f}{r}\Big(\fft{f'}{f} - \fft{h'}{h} + \ft12 r \phi'^2\Big)=0\,,\cr
E^{t}_t + E^{r}_r &=& \fft{f}{r}\Big(\frac{p^2 e^{a \phi }+q^2 e^{-a \phi }}{2 f r^3}+\frac{f'}{f}+\frac{2 (f-1)}{f r}+\frac{h'}{h}\Big)=0\,,\cr
E^{t}_t + E^{2}_2 &=& \ft12f \Big(\frac{f' h'}{2 f h}+\frac{f'}{f r}-\frac{2}{f r^2}+\frac{h''}{h}+\frac{3 h'}{h r}-\frac{h'^2}{2 h^2}+\frac{2}{r^2}
\Big)=0\,.
\eea
The last equation above can be integrated, yielding
\be
f=\frac{h \left(4 h r^2+\mu \right)}{r^2 \left(r h'+2 h\right)^2}\,,
\ee
where $\mu$ is an integration constant associated with non-extremality.  Setting $\mu=0$ yields the extremal solution with
\be
f^{-1}=\Big(1 + \fft{rh'}{2h}\Big)^2\,.\label{extcond}
\ee
(Dyonic solutions were also investigated in \cite{1504.07657} in a
different but equivalent parameterization of the metric. This included a
relation implied by extremality that is equivalent to our equation
(\ref{extcond}).)

\section{Special exact solutions}

In the previous section, we obtained the equations of the EMD theory for static and spherically-symmetric black holes carrying both electric and magnetic charges. The general solutions are unknown; however, many special classes of the solutions are known and we present them in this section.

\subsection{Purely electric or purely magnetic solutions}

The metrics of either purely electrically-charged or magnetically-charged black holes take the same form.
In the coordinate choice of \cite{Duff:1996hp,Cvetic:1996gq}, they are given by
\be
ds^2 = - H^{-\fft{2}{a^2+1}} \tilde f dt^2 + H^{\fft{2}{a^2+1}}\left(\fft{dr^2}{\tilde f}+ r^2d\Omega^2_2 \right) \,, \quad H = 1 + \fft{\mu s^2}{r}\,,\quad\tilde f= 1 - \fft{\mu}{r}\,,
\ee
with the matter fields given by
\bea
\hbox{electric}:&& F_1 = \fft{q}{r^2} H^{-2}dt \wedge dr \,, \qquad \phi = \fft{2a}{a^2+1} \log{H} \,, \qquad
q=\fft{2\mu}{\sqrt{a^2+1}} c\,s\,,\nn\\
\hbox{magnetic:}&& F = p\, \Omega_\2 \,, \qquad  \phi = -\fft{2a}{a^2+1} \log{H} \,, \qquad
p=\fft{2\mu}{\sqrt{a^2+1}} c\,s\,,
\eea
where $c=\cosh\delta$ and $s=\sinh\delta$. The horizon is located at $r_0=\mu$ and the extensive thermodynamic quantities are
\bea
\hbox{electric}:&& M = \ft12 \mu (1+\ft{2}{1+a^2}s^2) \,, \qquad Q = \ft14 q \,,\qquad S = \pi \mu^2\, c^{\fft{4}{1+a^2}} \,,\nn\\
\hbox{magnetic}:&& M = \ft12 \mu (1+\ft{2}{1+a^2}s^2) \,, \qquad P = \ft14 p \,,\qquad S = \pi \mu^2\, c^{\fft{4}{1+a^2}}\,.
\eea
From these, we can express the entropy in terms of mass and charges:
\bea
S(M,Q)&=&\fft{\pi  2^{5-\frac{N}{2}} (M^2-N Q^2)^{2-\frac{N}{2}} \Big(M \sqrt{M^2-\fft{8 (N-2) Q^2}{N}}+M^2-2 (N-2) Q^2\Big)^{\fft{N}2}}{(4-N) N M\sqrt{M^2-\fft{8 (N-2) Q^2}{N}}+((N-4) N+8)M^2-4 (N-2) N Q^2},\nn\\
S(M,P)&=& \frac{\pi  2^{5-\frac{N}{2}} (M^2-N P^2)^{2-\frac{N}{2}} \Big(M\sqrt{M^2-\frac{8 (N-2) P^2}{N}}+M^2-2 (N-2) P^2\Big)^{\fft{N}2}}{(4-N) N M\sqrt{M^2-\frac{8 (N-2) P^2}{N}}+((N-4) N+8)M^2-4 (N-2) N P^2}.\label{singlechargeentropy}
\eea
The extremal limit corresponds to taking $\mu\rightarrow 0$, and $\delta\rightarrow \infty$ while keeping $q$ or $p$ fixed. Under this limit, we have
\bea
\hbox{electric}:&& M_{\text{ext}} = \ft{2}{\sqrt{a^2+1}} Q=NQ \,, \qquad S_{\text{ext}} = 0 \,,\nn\\
\hbox{magnetic}:&& M_{\text{ext}} = \ft{2}{\sqrt{a^2+1}} P=NP \,, \qquad S_{\text{ext}} = 0 \,.
\eea
These extremal black holes with $N=2,3$ and 4 can be viewed as bound states with zero threshold energy
\cite{Rahmfeld:1995fm}.

\subsection{Known dyonic solutions}\label{knowndyonic}

The EMD theory admits a class of dyonic black holes with constant dilaton $\phi_0$.  The solution is given by
\be
h=f=1 -\fft{2M}{r} + \fft{q p}{2r^2}\,,\qquad
\phi=\phi_0\,,\qquad q=\sqrt2 q_0 e^{\fft12a \phi_0}\,,\qquad
p=\sqrt2 q_0 e^{-\fft12 a \phi_0}\,.
\ee
Note that this solution has constant $\phi$ and hence we can make a
constant shift of $\phi$ by $\phi_0$ and obtain the RN-like black hole
with equal electric and magnetic charges. In particular, in the
extremal limit, we have
\bea
ds^2 &=&  -\big(1-\fft{r_0}{r}\big)^2 dt^2 + \fft{dr^2}{\big(1-\fft{r_0}{r}\big)^2}+
r^2 d\Omega_2^2\,,\nn\\
\phi &=& 0\,,\qquad P=Q=\ft14\sqrt{2}r_0\,.\label{constantphisol}
\eea
The mass and charges are related by $M=2\sqrt2 Q=2\sqrt2 P$.

In this paper, however, we are interested in solutions with a running dilaton
such that $\phi\rightarrow 0$ asymptotically at infinity. Explicit
solutions are unknown for generic values of $a$. For the dilaton coupling
constant $a$ taking the values $a=0$, 1 or $\sqrt3$, exact solutions for
dyonic black holes are known. The $a=0$ case yields the RN dyonic black hole:
\bea
ds^2 &=& - f dt^2 + \fft{dr^2}{f} + r^2 d\Omega_2^2\,,\nn\\
F &=& \fft{q}{r^2} dt\wedge dr + p \Omega_\2\,,\qquad f = 1 - \fft{2M}{r} + \fft{q^2 + p^2}{4r^2}\,.
\eea
The solution has mass $M$ and electric and magnetic charges
$Q=\ft14 q$ and $P=\ft14 p$, respectively.  In the extremal limit, we have
\be
M=2\sqrt{P^2 + Q^2}\,.
\ee
To be precise, the dilaton is not running in this solution.

For $a=1$, the dyonic black solution is
\bea
ds^2 &=& - (H_1 H_2 )^{-1} {\tilde f}dt^2 + (H_1 H_2) \left( \fft{dr^2}{\tilde f} + r^2 d\Omega^2_{(2)} \right) \,,\nn\\
F &=& \fft{q}{r^2} H_1 ^{-2}\,dt \wedge dr + p\, \Omega_{(2)} \,, \qquad \phi = \log \fft{H_1}{H_2} \,,\nn\\
{\tilde f} &=& 1 - \fft{\mu}{r} \,,\qquad H_i= 1+ \fft{\mu s_i^2 }{r} \,,\qquad q= \sqrt{2} \mu s_1 c_1 \,, \quad p = \sqrt{2} \mu s_2 c_2 \,,
\eea
where $c_i=\cosh\delta_i$ and $s_i=\sinh\delta_i$.  The solution contains three independent integration constants $(\mu, \delta_1,\delta_2)$, parameterizing the mass, electric and magnetic charges
\be
M = \ft12 \mu (1+ s_1^2 + s_2^2) \,,
\qquad Q = \ft14 q \,, \qquad P= \ft14 p \,.
\ee
The solution describes a black hole whose horizon is
located at $r_0=\mu$, and the corresponding Bekenstein-Hawking entropy is
\be
S = \pi \mu^2 c_1^2 c_2^2 \,.
\ee
In the extremal limit, where $\mu\rightarrow 0$ and $\delta\rightarrow \infty$,
while keeping $(Q,P)$ fixed, the mass and entropy are related to $Q$ and $P$ by
\be
M = \sqrt{2} (Q + P )\,,\qquad  S = 8 \pi Q P \,.
\ee
It should be pointed out that the embedding of the $a=1$ EMD theory in supergravities requires $F\wedge F=0$. Thus the dyonic solution presented here is not a supergravity solution.

The $a=\sqrt3$ dyonic black hole was constructed in \cite{gibmae,gibwil}.
In this paper, we present the solution in a form where the
integration constants are $(\mu,\delta_1,\delta_2)$, as in the
previous examples. Following the notation of \cite{lupapo}
and making the reparameterization for the integration constants
\be
\lambda_1= \frac{s_1^2 (c_1^2 +1 ) c_2^2}{c_1^2 (s_1^2 s_2^2 +c_1^2+c_2^2)} \,,\qquad \lambda_2= \frac{s_2^2 (c_2^2 +1 ) c_1^2}{c_2^2 (s_1^2 s_2^2 +c_1^2+c_2^2)}\,,
\ee
we find that the solution becomes
\bea
ds^2 &=& - (H_1 H_2)^{-\fft12} {\tilde f} dt^2 + (H_1 H_2)^{\fft12}\left(\fft{dr^2}{\tilde f} +r^2 d\Omega_{(2)}^2 \right) \,,\nn\\
F &=& \fft{q}{r^2} H_1^{-2} H_2 dt \wedge dr + p\, \Omega_{(2)} \,,\qquad \phi = \fft{\sqrt{3}}{2} \log\fft{H_1}{H_2} \,, \nn\\
{\tilde f} &=& 1 - \fft{\mu}{r} \,,\qquad
H_i = 1+ \fft{\mu s^2_i}{r} + \fft{\mu^2 c_i^2 s^2_1 s^2_2}{2(c^2_1 + c^2_2)r^2}\,, \nn\\
&&q=\mu s_1 c_1 \sqrt{\ft{1+ c^2_1}{c^2_1 + c^2_2}} \,,\qquad p=\mu s_2 c_2 \sqrt{\ft{1+ c^2_2}{c^2_1 + c^2_2}}\,.
\eea
This solution describes a black hole whose horizon is located at $r_0=\mu$.
The mass, electric and magnetic charges, and entropy are given by
\be
M = \ft14\mu (c^2_1 + c^2_2) \,,\qquad Q = \ft14 q \,, \qquad P= \ft14 p \,,\qquad
S = \frac{\pi  c_1 c_2 (c_1^2 +1) (c_2^2+1)}{2 (c_1^2+c_2^2)}\mu^2 \,.
\ee
In this parameterization, $(\delta_1,\delta_2)$ are two independent constants.
The special cases of $\delta_2=0$ or $\delta_1=0$ lead to purely electric or
purely magnetic black holes, respectively.
The extremal limit corresponds to taking $\mu\rightarrow$ and $\delta_i\rightarrow \infty$, while keeping the charges $(Q,P)$ fixed.  Specifically, we introduce $\tilde s_i$,
\be
\mu = \sqrt{\tilde{s}_1^2 + \tilde{s}_2^2} \; \epsilon \,,\qquad s^2_i = \fft{4 \tilde{s_i}^2}{\epsilon} \,,
\ee
and let $\epsilon \rightarrow 0$, under which $c_i\rightarrow s_i$.  We obtain
\be
M = (\tilde{s}_1^2+\tilde{s}_2^2)^\ft32 \,, \qquad Q = \tilde{s}_1^3 \,, \qquad P = \tilde{s}_2^3 \,.
\ee
Thus the mass and entropy for the extremal dyonic black hole are
\be
M = (Q^\fft23 + P^\fft23)^\fft32 \,,\qquad S=8\pi Q P\,.
\ee

\section{Approximate solutions of extremal dyonic black holes}

In this section, we consider extremal dyonic black holes for the EMD theory
with a generic value of the dilaton coupling constant $a$.
The extremality condition is given by (\ref{extcond}), and
thus the solution takes the form
\bea
ds^2 &=& - h dt^2 + (h + \ft12 r h')^2 \fft{dr^2}{h^2} + r^2 d\Omega_2^2\,,\qquad \phi=\phi(r)\,,\nn\\
F &=& \fft{q}{r^2} e^{-a\phi} \sqrt{h} \big(1 + \fft{rh'}{2h}\big)\, dr\wedge dt + p\, \Omega_\2\,.
\eea
The remaining two functions $h$ and $\phi$ satisfy
\bea
\fft{h''}{h} + \fft{2h'}{r h} + \fft{r h'^3}{4 h^3} &=& \fft{(2h + r h')^3}{16 r^4 h^3} (p^2 e^{a\phi} +q^2 e^{-a\phi})\,,\nn\\
\phi'' - \ft14 \phi' \big( r\phi'^2 - \fft{4h'}{h} - \fft{8}{r}\big) &=& \fft{a(2h + r h')^2}{8 r^4 h^2}(p^2 e^{a\phi} - q^2 e^{-a\phi})\,,\label{extremaleom}
\eea
together with the first-order constraint
\be
\phi'^2 + \fft{h'^2}{h^2} = \fft{(2h + r h')^2}{4 r^4 h^2} (p^2 e^{a\phi} + q^2 e^{-a\phi})\,.\label{extremalcons}
\ee

\subsection{Approximate solutions}

Exact solutions for generic dilaton coupling $a$ are unknown except for
the case with a constant dilaton.  In particular, if we insist that
$\phi$ vanish asymptotically at infinity, then in the constant-dilaton case
it vanishes everywhere and
the corresponding solution is RN-like with $P=Q$.  In this section, we
consider solutions with $P\ne Q$, and hence with a running dilaton.
We introduce a small parameter $\epsilon$, characterising the deviation
away from $P=Q$, by writing
\be
Q=\ft14 q = Q_0 e^{\ft12 a \epsilon}\,,\qquad
P=\ft14 p = Q_0 e^{-\ft12 a \epsilon}\,,\qquad
Q_0=\fft{r_0}{2\sqrt2}\,.\label{approxcharge}
\ee
Note that this implies that we have
\be
S=8\pi P Q\,,
\ee
for extremal dyonic black holes. When $\epsilon=0$, we have $P=Q$ and
the exact solution is known, given by (\ref{constantphisol}).
In this section, we consider solutions with small $\epsilon$ as a
perturbation from (\ref{constantphisol}). Namely, we write the solutions
for $h$ and $\phi$ as
\be
h = \big(1 - \fft{r_0}{r}\big)^2 \Big[1 + \sum_{n\ge 1} \epsilon^n h_n(r)
\Big]\,,\qquad
\phi = \sum_{n\ge1} \epsilon^n \phi_n(r)\,.\label{hphiexpansion}
\ee
We solve the equations order by order in $\epsilon$ for $h_n$ and $\phi_n$.
We then find that at each order, the solutions can be obtained analytically,
and are given by
\bea
\fft{h}{x^2} &=& 1 + \epsilon^2 \frac{k\left(1-x^{2 k-1}\right)}{4(1-2 k)} +
\frac{\epsilon^4 k^2}{192 \left(16 k^2-1\right) (1-2 k)^2}\Big[2 \left(4 k^4+8 k^3+45 k^2+4 k-1\right)\cr
&&+3 (1-2 k)^2 \left(2 k^2-12 k-5\right) x^{4 k-1}-4 \left(8 k^4+10 k^3+57 k^2+5 k-5\right) x^{2 k-1}\cr
&&+3 (4 k-1) (4 k+1)^2 x^{4 k-2}\Big] + {\cal O}(\epsilon^6)\,,\nn\\
\phi &=& \epsilon (1-x^k) -\frac{\epsilon^3 k x^{k-1}}{24 \left(8 k^2-2 k-1\right)}\Big[\left(2 k^3-9 k^2+2\right) x^{2 k+1}+3 k (4 k+1) x^{2 k}\cr
&&-\left(2 k^3+3 k^2+3 k+2\right) x\Big] + {\cal O}(\epsilon^5)\,,
\eea
where we have expressed the dilaton coupling $a$ and the radial coordinate
$r$ in the form
\be
a^2 = \ft12 k (k+1)\,,\qquad x=1 - \fft{r_0}{r}\,.\label{asqandx}
\ee
The $x$ coordinate runs from 0 to 1, as the radial coordinate
$r$ runs from the horizon at $r=r_0$ to asymptotic infinity. Note
that the solutions have no branch cut singularities in the cases where
$k$ is taken to be an integer.  In deriving the above solutions,
we restricted the integration constants by imposing the following criteria:
(1) the quantity $h/x^2$ is regular on the horizon; (2) $h/x^2$ is set to
one as $x\rightarrow 1$; (3) $\phi$ is of order $\epsilon$ on the horizon
and vanishes asymptotically as $x\rightarrow 1$.  The solution is
then completely fixed at each order by these criteria.  In appendix A
we present the results for $(h/x^2)$ and $\phi$ up to and including
order $\epsilon^{10}$ and $\epsilon^{9}$, respectively.

It is clear that the solution is well defined from the horizon at $x=0$ to
asymptotic infinity at $x=1$, which describes an extremal black hole for each $k$.
The ADM mass is given by
\bea
M&=&r_0\Big[1 + \fft{\epsilon^2 k}{8} + \frac{\epsilon^4 k^2 \left(2 k^2+4 k-1\right)}{384 (4 k+1)}
\cr
&&+\frac{\epsilon^6 k^3 \left(24 k^5+64 k^4+52 k^3-18 k^2+34 k+19\right)}{46080 (4 k+1)^2 (6 k+1)}\cr
&& +\frac{\epsilon^8 k^4/10321920}{(4 k+1)^3 (6 k+1) (8 k+1)}\Big[384 k^8+2560 k^7+5792 k^6+4440 k^5\cr
&&+380 k^4-680 k^3-2532 k^2-1910 k-559]\cr
&&+\frac{\epsilon^{10} k^5/3715891200}{ (4 k+1)^4 (6 k+1)^2 (8 k+1) (10 k+1)}\Big[46080 k^{12}-435072 k^{11}\cr
&&-4154368 k^{10}-8803712 k^9-5889792 k^8+2243328 k^7+6981648 k^6\cr
&&+6048408 k^5+3652188 k^4+2110608 k^3+914808 k^2+288590 k+29161
\Big]\,,\nn\\
&&+ {\cal O}(\epsilon^{12})\,. \label{mass}
\eea
The electric and magnetic potentials $\Phi_q$ and $\Phi_p$ are presented
explicitly, as power series in $\epsilon$, in appendix \ref{app:approx}.
It can now be verified,
up to the order of $\epsilon^{10}$, that the first law of extremal
black hole thermodynamics is satisfied, namely
\be
dM=\Phi_q dQ + \Phi_p dP\,,\qquad M = \Phi_q Q + \Phi_p P\,.
\ee

\subsection{Mass-charge relation}

Having obtained the mass (\ref{mass}) and charges (\ref{approxcharge}) in
terms of $r_0$ and the small parameter $\epsilon$, we can test
the validity of the mass relation (\ref{rasheedconj}) that was
conjectured by Rasheed, order by order in powers of $\epsilon$.  We find that
\bea
M-\fft2{\sqrt{a^2 + 1}} (Q^b + P^b)^{\fft{1}{b}}=\ft18 k \Big(1 - \fft{(1+k) \log2}{\log(k(k+1)+2)}
\Big) r_0 \epsilon^2 + {\cal O}(\epsilon^4)\,.
\eea
Thus we see that unless $k=0,1,2$, corresponding to $a=0$, $a=1$ or
$a=\sqrt3$, the conjectured relation is violated at
the order $\epsilon^2$.

We have not succeeded in finding an exact mass-charge relation, based on the
data from our approximate solutions. However, we can improve on the
conjecture by Rasheed,  and thus we may consider the mass-charge relation
\be
M^2 - \ft{8}{k(k+1)+2} (P^2 + Q^2) -
\ft{k(k+1)}{k(k+1)+2} (8PQ)^{\fft{2}{k+1}} M^{\fft{2(k-1)}{k+1}}\approx
0\,.\label{approxrelation1}
\ee
As in the case of the Rasheed conjecture (\ref{rasheedconj}),
the above relation works exactly for
all $(P,Q)$ when $k=0,1,2$.  It also works for general $k$, in the
special cases of $P=0$, $Q=0$, or $P=Q$.  For $P\ne Q$, but with $\epsilon$
small, we find that the left-hand side of (\ref{approxrelation1}) is given by
\be
-\frac{k(k-2) (k-1) k^{\frac{1}{2} \left(k^2+k+6\right)} (k+1)^{\frac{1}{2} \left(k^2+k-2\right)}}{32 (4 k+1) \left(k^2+k+2\right)^{\frac{1}{2} k (k+1)}} r_0^{k (k+1)}\epsilon^4 + {\cal O}(\epsilon^6)\,.
\ee
Thus we see that the relation (\ref{approxrelation1})
captures the mass-charge relation up to order $\epsilon^4$, improving upon
the Rasheed conjecture (\ref{rasheedconj}) which already fails at order
$\epsilon^2$.

We can, of course, simply view the expression (\ref{mass}), which gives
the mass
as a power series in $\epsilon$, as a mass-charge relation that is valid up
to order $\epsilon^{10}$.  Thus, from (\ref{approxcharge}), we have that
\be
r_0=\sqrt{8PQ}\,,\qquad \epsilon= \fft{2}{a}\,
\arcsinh\Big(\fft{Q-P}{2\sqrt{PQ}}\Big)\,,
\ee
and so the right-hand side of (\ref{mass}) can be re-expressed as a
power series in the small quantity
\be
w=\fft{Q-P}{2\sqrt{(1+k)P Q}}\,.
\ee
The mass formula (\ref{mass}) therefore becomes
\bea
M &=& \sqrt{8PQ}\, \Big[ 1 + w^2 - \fft{1+2k+2k^2}{2(1+4k)}\, w^4
  + \fft{(1+2k+2k^2)(1+6k+16k^2+12k^3)}{2(1+4k)^2\, (1+6k)}
 \, w^6 \nn\\
&&-\fft{(1+2k+2k^2)(5+60k+382k^2+1316 k^3 + 2592 k^4 +
2560 k^5 + 960 k^6)}{8(1+4k)^3 (1+6k)(1+8k)}\, w^8  \label{mass2} \\
&&+ \fft{(1+2k+2k^2)}{8(1+4k)^4 (1+6k)^2 (1+8k) (1+10k)}\,
\Big(7+168k + 2030 k^2\nn\\
&&\qquad + 15644 k^3 + 80856 k^4 +
    288352 k^5 + 703808 k^6
 + 1147488 k^7\nn\\
&&\qquad + 1174336 k^8 + 670656 k^9
  + 161280 k^{10}\Big)\, w^{10}+\cdots\Big]\,,\nn
\eea
with the expansion on the right-hand side involving a function purely of
$P$ and $Q$.

It is noteworthy that in terms of the parameterization (\ref{approxcharge})
of $P$ and $Q$ in terms of $\epsilon$ and $r_0$, the exact mass relations
(\ref{relation1}) for the $a=0$, 1 and $\sqrt3$ extremal black holes become
\bea
k=0: &&\qquad a= 0 \,, \qquad M = 2 \sqrt{Q^2 + P^2}=r_0\,, \cr
k=1:&&\qquad a = 1 \,, \qquad M =
       \sqrt{2}(Q + P)=r_0\, \cosh\fft{\epsilon}{2}\,, \cr
k=2:&& \qquad a = \sqrt{3} \,, \quad M = (Q^\fft23 + P^\fft23)^\fft32
 = r_0\, \Big(\cosh\fft{\epsilon}{\sqrt3}\Big)^{3/2}\,.\label{relation10}
\eea
It is thus tempting to try an ansatz for the mass formula,
in the case of general $a$, of the
form
\be
\widetilde M= r_0\, (\cosh\alpha\epsilon)^\beta\,.\label{coshform}
\ee
The two constants $\alpha$ and $\beta$ can be chosen so as to match the
terms at orders $\epsilon^2$ and $\epsilon^4$ in the expansion (\ref{mass}),
implying that they are given by
\be
\alpha^2= \fft{k\,(2+4k-k^2)}{4(1+4k)}\,,\qquad
\beta= \fft{1+4k}{2+4k-k^2}\,.
\ee
Perhaps not surprisingly, the ansatz breaks down at order $\epsilon^6$,
with
\be
M-\widetilde M=
-\fft{(k-2)(k-1)k^4\, (k+1)(3k+2)\, r_0}{1920 (1+4k)^2 (1+6k)} \, \epsilon^6
+{\cal O}(\epsilon^8)\,.
\ee
Of course, this term and all subsequent terms vanish, as they must,
in the special cases $k=0$, $k=1$ and $k=2$ for which (\ref{coshform}) will
be exact.  Note that for general $k$, if (\ref{coshform}) is expressed
in terms of $P$ and $Q$, it becomes
\be
\widetilde M= 2^{\ft32 -\beta}\, (PQ)^{\ft12-\ft{\alpha\beta}{a} }\, \Big[
   Q^{\ft{2\alpha}{a} } + P^{\ft{2\alpha}{a} }\Big]^\beta\,,\label{coshconj}
\ee
which is a more general function than the rather natural-looking ansatz
(\ref{rasheedconj}) that was proposed by Rasheed.  (Only at $k=0$,
1 and 2 does (\ref{coshconj}) reduce to the form of the ansatz in
(\ref{rasheedconj}).)  And indeed, (\ref{coshconj}) gives an approximation
to the true mass formula (\ref{mass}) that works up to (but not including)
order $\epsilon^6$, whereas (\ref{rasheedconj}) works only up to
(but not including) order $\epsilon^4$.

   Another option for writing down a mass-charge relation is first to
invert the expansion (\ref{mass}) for $M$ as a power series in $\epsilon$,
re-casting it as an expansion for $\epsilon$ as a power series
in
\be
y\equiv \fft{M}{r_0} -1  = \fft{M}{\sqrt{8PQ}} -1\,.
\ee
Substituting this expansion into $\cosh a\epsilon$, which is equal
to $(P^2+Q^2)/(2PQ)$, we find
\crampest
\bea
&&\fft{(Q-P)^2}{2(k+1)PQ} = 2 y +
  \fft{(2k^2+2k+1)}{4k+1}\, y^2 -
  \fft{2k\, (k-1)(2k^2+2k+1)}{(4k+1)^2\, (6k+1)}\, y^3 \nn\\
&& +\fft{k\, (k-1)(2k^2+2k+1)(20 k^3+46 k^2+4k+5)}{
   2(4k+1)^3\, (6k+1)(8k+1)}\, y^4 \label{rel1}\\
&&
-\fft{3 k(k-1)(2k^2+2k+1)(144k^7+712 k^6 +904 k^5 + 230 k^4 +106 k^3
   -k^2 + 4k + 1)}{(4k+1)^4\, (6k+1)^2\ (8k+1)(10k+1)}\, y^5+
   \cdots\,.\nn
\eea
\uncramp
This provides a mass-charge equation relating $M$, $P$ and $Q$ that is
valid up to, but not including, order $\epsilon^{12}$.  Although this
form (\ref{rel1}) of the mass-charge relation is somewhat less convenient than
(\ref{mass2}), which directly gives an expression for $M$ as a function of
$P$ and $Q$, it does seem that the coefficients in the expansion are
rather simpler in (\ref{rel1}).

\begin{figure}[ht]
\begin{center}
\includegraphics[width=10cm]{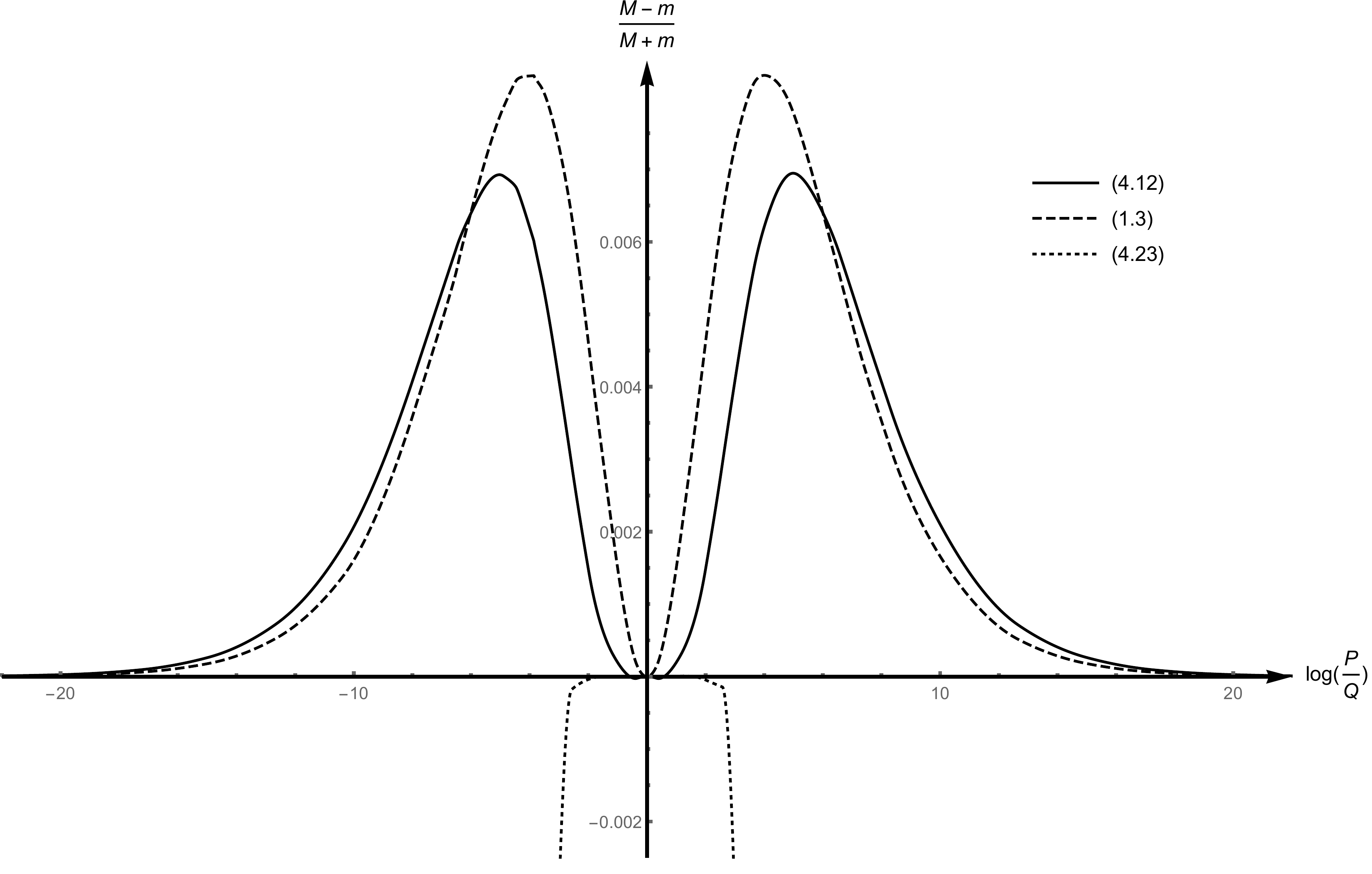}
\caption{This plot shows the difference between the approximate mass formula
$M$ and the actual mass $m$ obtained from the numerical calculations, with
horizontal axis $x=\log(P/Q)$ and vertical axis $y=(M-m)/(M+m)$.
In this plot, we fix $PQ = 1/8$, corresponding to a horizon radius
$r_0=1$. The dashed line is from the Rasheed formula,
the continuous line is for our new formula (\ref{approxrelation1}); the
dotted line for (\ref{rel1}). Thus for $|\log(P/Q)|<2$, the mass
formula (\ref{rel1}) is much better, but it quickly diverges from the
true mass as $|\log(P/Q)|$ increases, with a radius of convergence of
about 2.  Our new formula (\ref{approxrelation1}) is better than Rasheed's
at small $x$, but is not as good as $x$ increases.  However, overall
our expression (\ref{approxrelation1}) appears to improve
upon Rasheed's in fitting
the actual data.  The maximum error in our formula is about 0.6\%, whilst
Rasheed's is about 0.9\%. We also obtained similar results
for the case $a=\sqrt{10}$.}
\label{comparemassformulae}
\end{center}
\end{figure}

   We can use numerical methods in order to compare the accuracy of
the various approximate mass formulae in the cases where $Q$ and $P$ are
no longer close to one another.  Our numerical approach for constructing
the extremal dyonic black holes is discussed in Appendix \ref{app:numeric}.
Focusing on the case of $k=3$ (i.e.~$a=\sqrt6$) as an example,
we compare the three mass formulae (\ref{rasheedconj}),
(\ref{approxrelation1}) and (\ref{rel1}) with the actual mass obtained
from the numerical calculations. The results are plotted in
Fig.~\ref{comparemassformulae}. In this plot we fix $PQ=1/8$, so
that the horizon radius is $r_0=1$.  We plot the difference
$(M-m)/(M+m)$ as a function of $\log(P/Q)$ where $M$ is obtained from the
approximate mass formulae and $m$ is obtained from the numerical calculations.  We also obtained the analogous result for the $a=\sqrt{10}$ case.  We can see
from the plots that although Rasheed's mass formula and our mass formula
(\ref{approxrelation1}) are approximate, they both capture the mass-charge
relation quite accurately.

\section{Entropy super-additivity and sub-additivity}

   As we discussed in the introduction, entropy super-additivity is a
property exhibited by many black holes, although the dyonic black hole
in the $a=\sqrt3$ EMD theory provides a counter-example \cite{cvgiluponeg}.
Since the dyonic black holes in $a=0$ and $a=1$ EMD theory do, on the
other hand, obey the super-additivity property, it is therefore of interest
to study for general values of the dilaton coupling $a$.

   For a static black hole with mass $M$ and electric and magnetic charges $Q$
and $P$,  the entropy must be a function of these quantities,
\be
S=S(M,Q,P)\,.
\ee
We now consider the splitting of the black hole into two black holes
with masses and charges $(M_1,Q_1,P_1)$ and $(M_2,Q_2,P_2)$, with
\be
M=M_1 + M_2\,,\qquad Q=Q_1 + Q_2\,,\qquad P=P_1+P_2\,,
\ee
and we consider the quantity
\be
\Delta S \equiv S(M,Q,P) - S(M_1,Q_1,P_1) -S(M_2,Q_2,P_2)\,.
\ee
The splitting of the black hole is the inverse of the joining of
the two black holes, and the entropy is super-additive if
$\Delta S>0$, and sub-additive if $\Delta S<0$.

    Our goal in this section is to study $\Delta S$ in cases where a
non-trivial dyonic black hole with $Q\ne P$ is involved, and in the
earlier sections we have obtained approximate results for such black holes
in the case of the extremal limit.  Thus here,
we shall investigate the entropy additivity property for the splitting
of an extremal dyonic black hole with mass $M$ and electric and magnetic charges
$(Q,P)$ into two black holes where one is purely electric, with mass $M_1$
and charge $(Q,0)$,
 and the other is purely magnetic, with mass $M_2$ and charge $(0,P)$.
These purely electric and magnetic black holes are not necessarily
extremal.
Since we require $M=M_1+M_2$, and since $M_1$ and $M_2$ are bounded below
by the masses $M_1(Q,0)_{\rm ext}$ and $M_2(0,P)_{\rm ext}$
of the extremal black holes with these charges,
\be
M_1(Q,0) \ge M_1(Q,0)_{\rm ext}\,,\qquad
M_2(0,P) \ge M_2(0,P)_{\rm ext}\,,
\ee
it follows that we must have
\be
M(Q,P)_{\rm ext} \ge M(Q,0)_{\rm ext} + M(0,P)_{\rm ext}\,.\label{bombeqn}
\ee
The dyonic black holes satisfying this condition have been referred to
as ``black hole bombs'' \cite{gibkal,lupofission}. They are closely related to the 
$SL(3,R)$-Toda system and can be generalized to $SL(n,R)$-Toda black holes \cite{Lu:2013uia}.

   In the following subsections, we shall study $\Delta S$ as a function of
the dilaton coupling $a$ in two situations; in one, $Q$ and $P$ will be held
fixed, whilst in the other we shall hold the entropy $S=8\pi PQ$ of the
extremal dyonic black hole fixed, and look at $\Delta S$ as the ratio $P/Q$
is allowed to vary.

\subsection{Fixed $P$ and $Q$}

\begin{figure}[ht]
\begin{center}
\ \ \ \ \ \includegraphics[width=7cm]{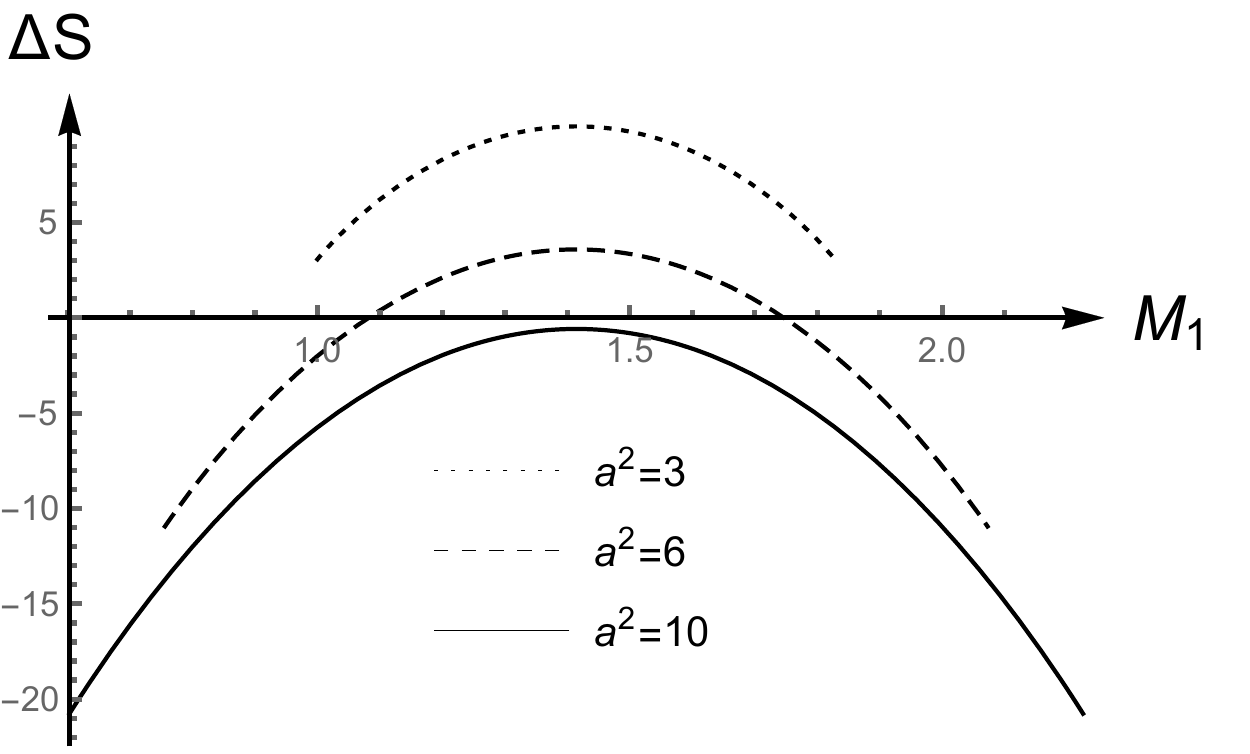}\ \
\includegraphics[width=7cm]{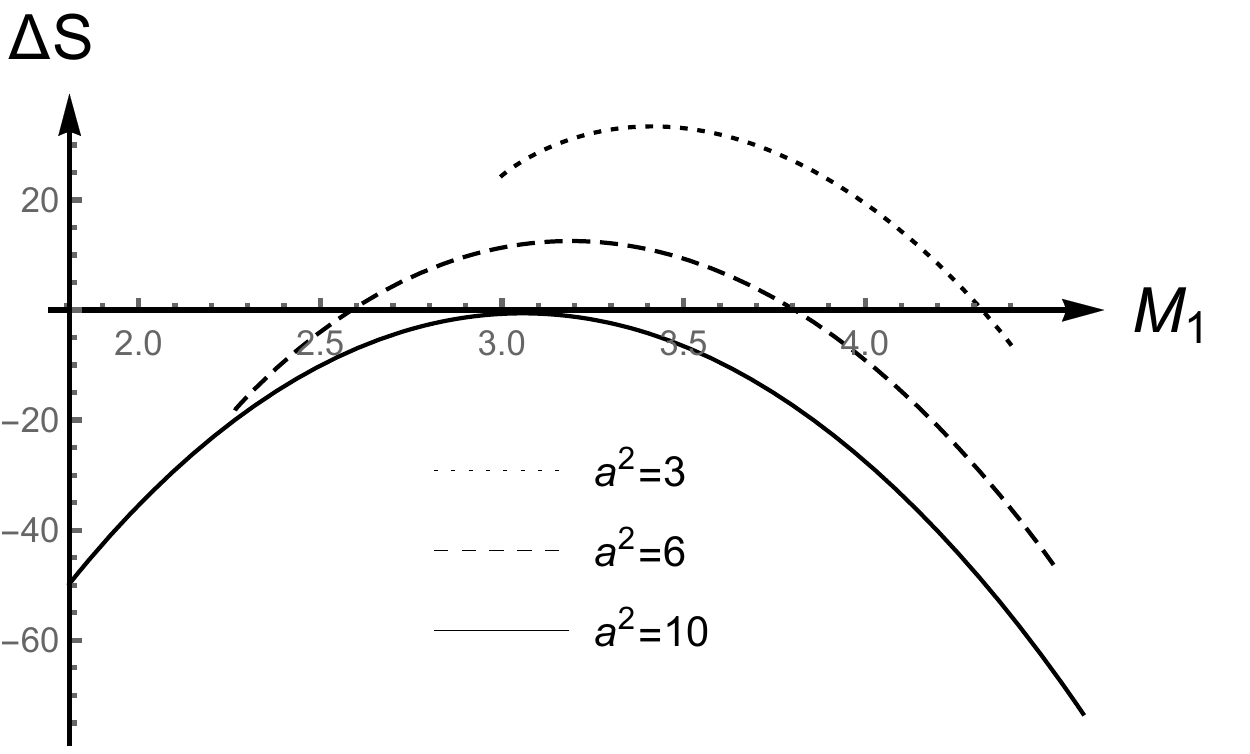}
\caption{In both plots, we present $\Delta S$ as a function of $M_1$,
which runs from $M_1^{\rm ext}$ to $M-M_2^{\rm ext}$, for the cases
$a^2=3$, 6 and 10.  In the left-hand plot we have chosen $Q=P=1$, and in
the right-hand plot $Q=3, P=1$.}
\label{fixedQPDeltaS}
\end{center}
\end{figure}

  We shall first study the simple case where $Q=P$, for which the mass
of the extremal dyonic black hole is known exactly
(see section \ref{knowndyonic}, under eqn (\ref{constantphisol})):
\be
M=M(Q,Q)_{\rm ext} =2\sqrt2 Q\,,\qquad S(Q,Q)_{\rm ext} =8\pi Q^2\,.
\ee
The masses $M_1=M_1(Q,0)$ and $M_2=M_2(0,Q)$ of the purely electric and
purely magnetic
black holes must satisfy $M_1 + M_2 = M=2\sqrt2 Q$.  Thus $M_1$ must satisfy
\be
\fft2{\sqrt{a^2+ 1}} Q=M_1^{\rm ext} \le M_1 =M-M_2
  \le M- M_2^{\rm ext}=  2\sqrt 2 Q - \fft2{\sqrt{a^2 + 1}} Q\,.
\ee
Note that from the inequality between the left-most and right-most terms in
this expression, we can immediately deduce that we must have $a\ge 1$.
The entropy of the two electric and magnetically charged black holes are
given in (\ref{singlechargeentropy}).  We can thus evaluate
$\Delta S$ as a function of $M_1$, for various values of the
dilaton coupling constant $a$.  Setting $Q=1$ without loss of generality,
these results are plotted in the left-hand graph in Fig.~\ref{fixedQPDeltaS},
for the examples of $a^2=3$, 6 and 10.

  We now turn to an example where $Q$ and $P$ are unequal.  For generic values
of $a$, this means that we must resort to numerical methods to
construct the extremal dyonic solution.  We shall consider the
example where $Q=3$ and $P=1$.  In this case, the
purely electric and magnetic solutions are
\be
M_1^{\rm ext} = \fft{6}{\sqrt{a^2+1}}\,,\qquad
M_2^{\rm ext}=\fft{2}{\sqrt{a^2+1}}\,.
\ee
The masses of the dyonic solutions for $a^2=6,10$ cases have to be obtained by numerical methods, described in appendix \ref{app:numeric}.  We find
\bea
a^2=3:&& M=(1 + 3^{\fft23})^{\fft32}\sim 5.41\,,\nn\\
a^2=6:&& M\sim 5.28\,,\nn\\
a^2=10:&& M\sim 5.20\,.
\eea
With these data, we can calculate $\Delta S$ as a function of
$M_1$, which runs from $M_1^{\rm ext}$ to $M-M_2^{\rm ext}$.
The result is presented in the right-hand graph in Fig.~\ref{fixedQPDeltaS}.

   The general picture that emerges from these examples is that in this region
of parameter space the entropy is super-additive for small values of $a^2$,
but it tends to become sub-additive for larger values of $a^2$.

\subsection{Fixed $S_{P,Q}=8\pi P Q$}

\begin{figure}[ht]
\begin{center}
\includegraphics[width=9cm]{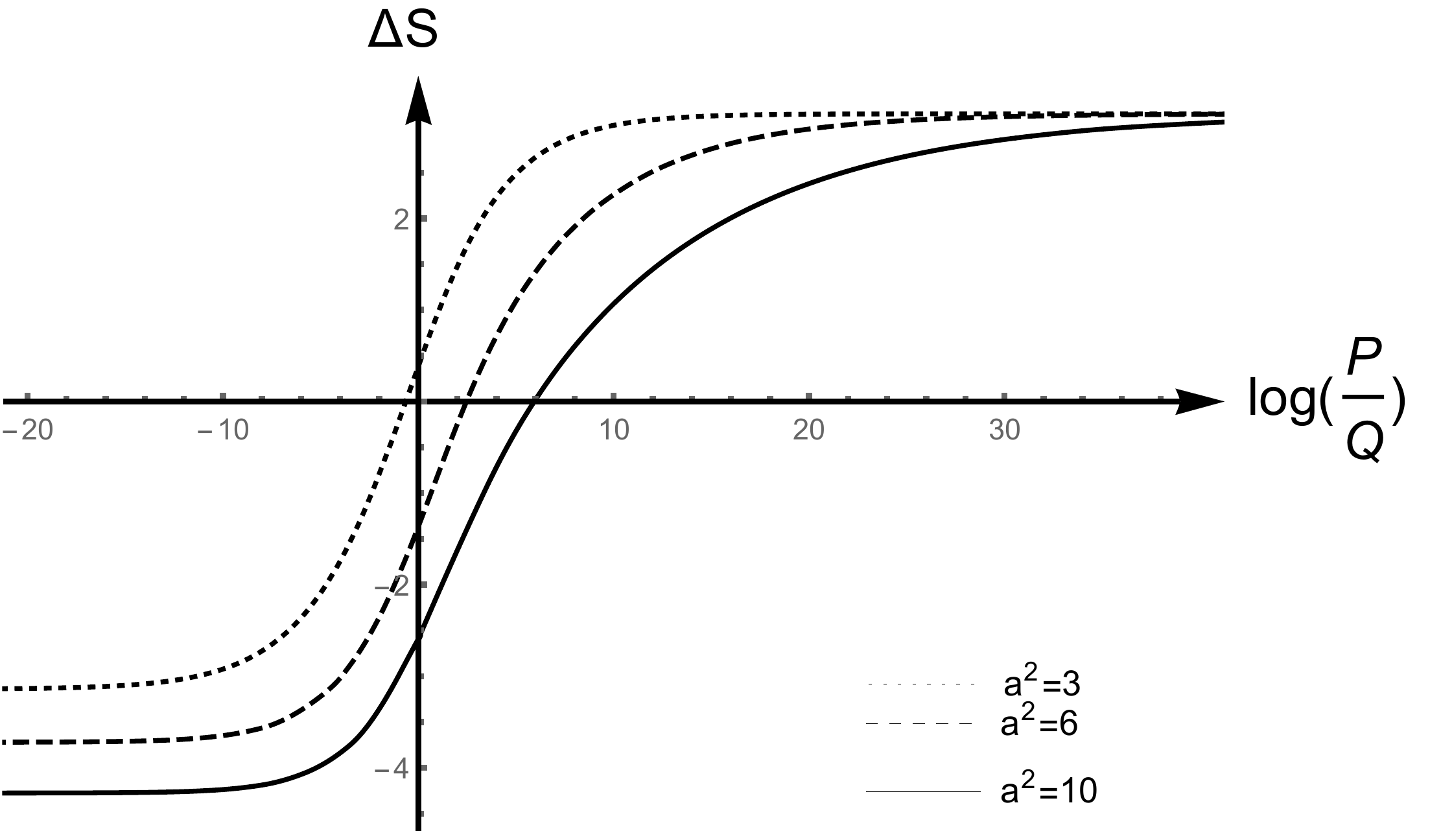}
\caption{The entropy difference for splitting a dyonic black hole to an
extremal magnetic black hole and a non-extremal electric black hole. The
electric and magnetic charges satisfy $QP=1/8$. The entropy is super-additive
for sufficiently large $P/Q$ and becomes sub-additive as $P/Q$ decreases.
Note that for all the $a^2=3$, 6 and 10 cases, the extremal dyonic black
hole can be viewed as a black hole bomb, satisfying
(\ref{bombeqn}).}
\label{fixedSPQ}
\end{center}
\end{figure}

   Here, we study the super-additivity of the entropy for the
splitting an extremal dyonic black hole into an extremal
magnetically-charged black hole and a non-extremal electrically-charged
black hole.  We look at $\Delta S$ as a function of the ratio $P/Q$
while holding the entropy of the extremal dyonic black hole fixed.
In particular we shall consider a horizon radius $r_0=1$ for the
extremal dyonic black hole.  In other words, it will have
$S=\pi$ and $PQ=1/8$.  Since the entropy $S_{\rm P}^{\rm ext}$ for
the extremal magnetically-charged black hole vanishes, we shall have
\be
\Delta S= S_{\rm P,Q}^{\rm ext} - S_Q^{\rm non-ext}\,.
\ee
We obtain and plot the results, again for the cases $a^2=3$, 6 and 10,
in Fig.~\ref{fixedSPQ}. The result for $a^2=3$ can be obtained analytically.
For the cases $a^2=6$ and $a^2=10$, we have used the numerical solutions
to find the mass.

First, we note that for large $P/Q$, the quantity $\Delta S$ is
positive and hence the entropy is super-additive. As $P/Q$ decreases,
$\Delta S$ becomes negative, and so the entropy becomes sub-additive.
Interestingly, in the limit $P/Q\rightarrow 0$, the quantity
$\Delta S$ approaches a negative constant that depends on $a$.
Assuming this conclusion is in general true for the $a^2\ge 1$ cases,
we can derive some useful information about the mass of the dyonic
black hole as $P/Q\rightarrow 0$.  It is evident that at leading order,
the mass in this limit is given by $\sqrt{N} Q$.  We postulate that
the sub-leading order is given by
\be
M=\sqrt{N} Q \Big(1 + \gamma\, \big(\fft{P}{Q}\big)^\delta + \cdots \Big)\,,\qquad \hbox{with}\qquad
\fft{P}{Q}\rightarrow 0\,.\label{nearlyelectric}
\ee
Here $(\delta,\gamma)$ are constants that at present we are unable
to determine for the case of a generic dilaton
coupling constant $a$.  After the dyonic black hole is split into purely
electric and purely magnetic black holes, only the non-extremal electric
black hole has an entropy. The mass of the electric black hole
is
\be
M_Q= \sqrt{N} Q \Big(1 + \gamma\, \big(\fft{P}{Q}\big)^\delta - \fft{P}{Q} + \cdots \Big)\,.
\ee
Assuming that $\delta <1$, it then follows from (\ref{singlechargeentropy})
that at leading order the entropy is given by
\be
S_{Q}^{\rm non-ext}=\pi (1 + a^2) 2^{2\fft{1+2a^2}{1+a^2}} a^{-\fft{4a^2}{1+a^2}} \big(\gamma \fft{P}{Q}\big)^{\fft{2a^2}{1+a^2}} Q^2 + \cdots\,.
\ee
Since $P Q$ is fixed, for this leading-order quantity to be constant
we must have
\be
\delta = \fft{1+a^2}{2a^2}\,.\label{deltavalue}
\ee
Note that $\delta \le 1$ for $a^2\ge 1$, thus the result is applicable
for $a^2>1$.  Then, at leading order as $P/Q\rightarrow 0$, $\Delta S$ is
given by
\be
\Delta S \rightarrow 8\pi PQ (1 - c)\,,\qquad c = (1 + a^2) 2^{\fft{a^2-1}{a^2+1}} a^{-\fft{4a^2}{a^2+1}}\gamma ^{\fft{2a^2}{a^2+1}}\,.
\ee
For the special case $a^2=3$, we have $\delta=2/3$ and $\gamma=3/2$, and
hence $c=2$. This is consistent with the $a=\sqrt3$ result in Fig.~\ref{fixedSPQ}.  For general $a>1$, we have determined $\delta$ by this method,
but the coefficient $\gamma$ cannot be determined generically.
The plots indicates that the constant $c$ is bigger than 1 and it
increases as $a$ increases.

\section{Conclusions}

   In this paper, we have studied the extremal static dyonic solutions of
the four-dimensional Einstein-Maxwell-Dilaton theory for generic
values of the dilaton coupling $a$.  Except in the special cases $a=0$,
$a=1$ and $a=\sqrt3$, where the explicit solutions can be constructed, it
is necessary to use approximation techniques or numerical methods in order
to study the solutions.

   Our method for constructing approximate solutions is based on the
observation that one can solve the equations explicitly, for any value
of $a$, in the special case when the electric charge $Q$ and the magnetic
charge $P$ are equal.  (The dilaton becomes constant in this case, and
the solution is just an extremal Reissner-Nordstr\"om $Q=P$ dyonic
black hole.)  We then construct the solution, in terms of series expansions
in powers of the small parameter $\epsilon = a^{-1}\, \log(Q/P)$, in the
regime where the ratio $Q/P$ is close to unity.

   One of our reasons for studying this problem is that there is a
long-standing conjecture by Rasheed \cite{rasheed}, dating back to 1995,
for a possible expression for the mass of the extremal dyonic black
hole as a function of the electric and magnetic charges and the dilaton
coupling $a$.  The power-series solutions that we obtain for $Q/P$ close to
unity are sufficient to be able to demonstrate that the conjectured
mass formula in \cite{rasheed} fails, at order $\epsilon^2$,
when $a$ is not equal to one of the
three values mentioned above for which the exact dyonic solutions are known.
We were able to show, however, by means of a numerical construction
of the dyonic solutions, that the conjectured mass formula is a reasonably
good approximation even as one moves away from the $Q\sim P$ regime of
our power-series solutions.
We have also made some somewhat improved conjectures for mass formulae, which
work up to higher orders in $\epsilon$, but again, they are only
approximations.  However, our findings from the numerical construction of
the dyonic black hole solutions suggest that both the Rasheed formula
(\ref{rasheedconj}) and our formula (\ref{approxrelation1}) capture the
essence of the mass-charge relation for extremal dyonic black holes
quite accurately, even when $Q/P$ is far away from unity.

   Our perturbative construction of the extremal dyonic black hole
solutions made use of the fact that we know the exact solution, for all
$a$, when $Q=P$, and so we could expand around this solution in terms
of the small parameter $\epsilon = a^{-1}\, \log(Q/P)$.  Of course, we also
know the exact solution when $P=0$ or $Q=0$, and one might wonder whether
this could provide another starting point for obtaining a solution
as a perturbative expansion valid for very large $P/Q$ or $Q/P$.
However, this does not seem to be a promising option.  The problem is that
no matter how large or small $Q/P$ is, as long as $Q$ and $P$ are both
non-zero the near-horizon geometry of the extremal black hole is of the
form AdS$_2\times S^2$.  By contrast, the horizon is singular for an
extremal black hole carrying purely electric or purely magnetic charge
(except in the case $a=0$).  Thus the starting point for such a
perturbative expansion would be singular.  It is interesting, nevertheless,
that our numerical construction of the black hole solutions led us to
a mass formula of the form (\ref{nearlyelectric}) for a nearly electric
extremal black hole, with $\delta$ given by (\ref{deltavalue}).

The other main topic in this paper has been the study of the circumstances
under which the property of super-additivity of the entropy of the dyonic
black holes is satisfied.  This property, defined in our discussion in the
introduction, has been found to be satisfied by many classes of black hole
solutions.  But already, in \cite{cvgiluponeg}, it was observed that
it could be violated in certain regimes for dyonic $a=\sqrt3$ EMD black
holes.   Our findings in this paper extend this observation further, and
indicate that entropy super-additivity can be violated in certain parameter
regimes for all the dyonic EMD black holes with larger values of $a$.

 All
the examples where entropy super-additivity fails are associated with the
situation where a ``black-hole bomb'' can occur, of the kind discussed
for $a=\sqrt3$ extremal dyonic black holes in \cite{gibkal,lupofission}.
Namely, these are situations where the mass of an extremal dyonic black hole
is greater than the sums of the masses of separated pure electric and
pure magnetic black holes of the same total charge, as in (\ref{bombeqn}).
This can happen, as we saw, when the dilaton coupling $a$ is such that
$a^2>1$.

   In \cite{cvgiluponeg}, the relation between Hawking's black hole
area theorem and entropy super-additivity was discussed.  The area theorem
asserts that, subject to cosmic censorship,
the area (or entropy) of the horizon of a black hole final state
formed from the coalescence of two black holes cannot be less than the sum of the
two original horizon areas (or entropies).  The area theorem would then imply
entropy super-additivity \cite{cvgiluponeg}.
However, an important distinction is that the
area theorem is only applicable in cases where the coalescence of the
black holes is physically possible.  As discussed in \cite{larsen} for the
$a=\sqrt3$ dyonic black hole, if a static dyonic black hole separated into
a pure electric and a pure magnetic black hole, there would be an angular
momentum proportional to $Q P$ and so the decay to purely static constituents
would seem to be ruled out by angular momentum conservation.  Thus the
breakdown of entropy super-additivity that we have seen in cases where
$a^2>1$ in this paper need not be in conflict with the Hawking area theorem.

  An intriguing point, worthy of further investigation, is that
one can seemingly sidestep the complication of angular momentum in the
electromagnetic field of a decomposed dyonic black hole by considering a
slight modification of the original EMD theory (\ref{EMD}), in which a second
electromagnetic field is introduced, with the Lagrangian
\be
{\cal L}=\sqrt{-g} (R - \ft12 (\partial\phi)^2 -
\ft14 e^{a\phi} F_1^2 -\ft14 e^{-a \phi} F_2^2\,.
\ee
This theory admits black hole solutions where $F_1$ and $F_2$ carry
electric charges $Q_1$ and $Q_2$ that are essentially identical to the
dyonic solutions of the EMD theory (\ref{EMD}), where the EMD charges are
$Q=Q_1$ and $P=Q_2$ \cite{hong}.  One then has all the same features
of the occurrence of black-hole bomb solutions for $a^2>1$, and violations
of entropy super-additivity, as we saw earlier in the EMD theory.  Now,
however, there is no angular momentum in the electromagnetic field(s) for
separated component black holes, and so the argument in \cite{larsen}
that could have ruled out the dyon decay process no longer applies.  It would
be interesting to investigate this further, to see how the occurrence of
entropy sub-additivity could be compatible with the Hawking area theorem
for the black holes in this theory.

\section*{Acknowledgments}

We thank Gary Gibbons for helpful discussions.
W.-J.G.~and H.L.~are supported in part by NSFC grants No.~11475024 and No.~11875200.
C.N.P.~is supported in part by DOE grant DE-FG02-13ER42020.

\appendix

\section{Approximate solution up to $\epsilon^{10}$ order}
\label{app:approx}

In this appendix, we present the dyonic extremal black hole solution (\ref{hphiexpansion}) up to and including the order $\epsilon^{10}$.  The results are necessary to derive the mass formula (\ref{mass}). Having defined (\ref{asqandx}), we have
\bea
\fft{h}{x^2} &=& 1 + \epsilon^2 \frac{k\left(x-x^{2 k}\right)}{4(1-2 k) x} +
\frac{\epsilon^4 k^2}{192 \left(16 k^2-1\right) (1-2 k)^2 x^2}\Big(2 \left(4 k^4+8 k^3+45 k^2+4 k-1\right) x^2\cr
&&+3 (1-2 k)^2 \left(2 k^2-12 k-5\right) x^{4 k+1}-4 \left(8 k^4+10 k^3+57 k^2+5 k-5\right) x^{2 k+1}\cr
&&+3 (4 k-1) (4 k+1)^2 x^{4 k}\Big)\nn\\
&&-\frac{\epsilon^6 k^3}{23040 (2 k-1)^3 (4 k-1) (4 k+1)^2 (6 k-1) (6 k+1) x^3}\Big[\cr
&&-45 (2 k-1)^2 (4 k+1) (6 k-1) (6 k+1)^2 \left(2 k^2-12 k-5\right) x^{6 k+1}\cr
&&+30 (4 k-1) (4 k+1)^2 (6 k-1) (6 k+1) \left(4 k^3+6 k^2+18 k+7\right) x^{4 k+1}\cr
&&+30 (2 k-1)^2 (6 k-1) (6 k+1) \left(2 k^2-12 k-5\right) \left(4 k^3+6 k^2+18 k+7\right) x^{4 k+2}\cr
&&-45 (2 k-1)^3 (4 k-1) \left(12 k^5-168 k^4+384 k^3+478 k^2+154 k+15\right) x^{6 k+2}\cr
&&-4 (6 k-1) \Big(816 k^8+2596 k^7+10976 k^6+17565 k^5+30794 k^4\cr
&&+16852 k^3+1980 k^2-638 k-91\Big) x^{2 k+2}+4 \Big(576 k^9+2880 k^8\cr
&&+10040 k^7+18424 k^6+35874 k^5+16588 k^4+4178 k^3+657 k^2+32 k+1\Big) x^3\cr
&&90 k (4 k-1) (4 k+1)^2 (6 k-1) (6 k+1)^2 x^{6 k}\Big]\nn\\
&& +\frac{\epsilon^8 k^4}{5160960 (1-4 k)^2 (4 k+1)^3 \left(2304 k^4-100 k^2+1\right) (1-2 k)^4x^4}\Big[\cr
&&840 k (4 k-1)^2 (4 k+1)^3 (6 k-1) (6 k+1) (8 k-1)^2 (8 k+1)^2 x^{8 k}\cr
&&-32 (4 k-1) (8 k-1) \Big(142848 k^{13}+739008 k^{12}+3101104 k^{11}+7958032 k^{10}\cr
&&+16656424 k^9+23696532 k^8+26928519 k^7+18001947 k^6+6125349 k^5\cr
&&+688362 k^4-88674 k^3-2111 k^2+2455 k+205\Big) x^{2 k+3}\cr
&&+504 (4 k-1) (4 k+1)^2 (6 k-1) (8 k-1) (8 k+1) \Big(288 k^8+888 k^7+2808 k^6\cr
&&+4150 k^5+5382 k^4+3066 k^3+395 k^2-154 k-23\Big) x^{4 k+2}\cr
&&+504 (2 k-1)^2 (6 k-1) (8 k-1) (8 k+1) \left(2 k^2-12 k-5\right) \Big(288 k^8+888 k^7\cr
&&+2808 k^6+4150 k^5+5382 k^4+3066 k^3+395 k^2-154 k-23\Big) x^{4 k+3}\cr
&&-5040 k (2 k+1) (4 k-1)^2 (4 k+1)^2 (6 k-1) (6 k+1)^2 (8 k-1) (8 k+1)\cr
&&\times\left(k^2+k+3\right) x^{6 k+1}-2520 (2 k-1)^2 (2 k+1) (4 k-1) (4 k+1) (6 k-1)\cr
&&\times(6 k+1)^2 (8 k-1) (8 k+1) \left(k^2+k+3\right) \left(2 k^2-12 k-5\right) x^{6 k+2}\cr
&&-2520 (2 k-1)^3 (2 k+1) (4 k-1)^2 (8 k-1) (8 k+1) \left(k^2+k+3\right)\cr
 &&\times \left(12 k^5-168 k^4+384 k^3+478 k^2+154 k+15\right) x^{6 k+3}\cr
 &&+5040 k (2 k-1)^2 (4 k-1) (4 k+1)^2 (6 k-1) (6 k+1) (8 k-1) (8 k+1)^2\cr
  &&\times\left(2 k^2-12 k-5\right) x^{8 k+1}+315 (2 k-1)^3 (4 k+1) (8 k-1) (8 k+1)^2\cr
   &&\times \left(1056 k^7-14736 k^6+40648 k^5+21908 k^4-6720 k^3-3072 k^2+206 k+85\right) x^{8 k+2}
\cr
&&+105 (2 k-1)^4 (4 k-1)^2 (4 k+1) (6 k-1) \Big(288 k^7-6816 k^6+41288 k^5-42780 k^4\cr
&&-98488 k^3-49044 k^2-9238 k-585\Big) x^{8 k+3}+8 \Big(1253376 k^{15}+10708992 k^{14}\cr
&&+36770816 k^{13}+94049792 k^{12}+188965248 k^{11}+274829056 k^{10}+333916576 k^9\cr
&&+198513792 k^8+76889376 k^7+16710012 k^6-958500 k^5-1023835 k^4-208600 k^3\cr
&&-25410 k^2-692 k+1\Big) x^4
\Big]\nn\\
&&+\frac{\epsilon^{10} k^5/1857945600}{(1-4 k)^2 (2 k-1)^5 (4 k+1)^4 (6 k-1) (6 k+1)^2 ((8 k)^2-1)((10 k)^2-1) x^5}\Big[\cr
&&18900 k (4 k+1)^4 (5 k-1) (6 k-1) (6 k+1)^2 (8 k-1) (8 k+1) (10 k+1)^2 (1-4 k)^2 \cr
&&\times(1-10 k)^2 x^{10 k}+16 (10 k-1) \Big(24454103040 k^{20}+180194770944 k^{19}\cr
&&+818173599744 k^{18}+2620475621888 k^{17}+6424614622720 k^{16}+12257466595232 k^{15}\cr
&&+18895354632528 k^{14}+22768387442448 k^{13}+21525540388776 k^{12}\cr
&&+14557877561170 k^{11}+6343842768089 k^{10}+1440749945746 k^9-16416399844 k^8\cr
&&-81249938138 k^7-13005568630 k^6-2134737402 k^5-422081208 k^4-38850528 k^3\cr
&&-1452596 k^2+118640 k+7381\Big) x^{2 k+4}-240 (4 k-1) (4 k+1)^2 (6 k+1)\cr
 &&\times (8 k-1) (10 k-1) (10 k+1) \Big(1953792 k^{13}+9522432 k^{12}+33807040 k^{11}\cr
 &&+77129152 k^{10}+137796448 k^9+176370384 k^8+170812668 k^7+107973084 k^6\cr
 &&+35323968 k^5+1927524 k^4-1522824 k^3-84773 k^2+27358 k+2497\Big) x^{4 k+3}\cr
&&-240 (1-2 k)^2 (6 k+1) (8 k-1) (10 k-1) (10 k+1) \left(2 k^2-12 k-5\right) \Big(1953792 k^{13}\cr
&&+9522432 k^{12}+33807040 k^{11}+77129152 k^{10}+137796448 k^9+176370384 k^8\cr
&&+170812668 k^7+107973084 k^6+35323968 k^5+1927524 k^4-1522824 k^3\cr
&&-84773 k^2+27358 k+2497\Big) x^{4 k+4}+15120 k (4 k-1) (4 k+1)^2 (6 k-1)\cr
 &&\times (6 k+1)^2 (8 k-1) (8 k+1) (10 k-1) (10 k+1) \Big(1776 k^8+5396 k^7+14776 k^6\cr
 &&+20945 k^5+23464 k^4+13512 k^3+1880 k^2-778 k-121\Big) x^{6 k+2}\cr
 &&+7560 (4 k+1) (6 k-1) (6 k+1)^2 (8 k-1) (8 k+1) (10 k-1) (10 k+1) (1-2 k)^2\cr
&&\times\left(2 k^2-12 k-5\right) \Big(1776 k^8+5396 k^7+14776 k^6+20945 k^5+23464 k^4+13512 k^3\cr
&&+1880 k^2-778 k-121\Big) x^{6 k+3}+7560 (2 k-1)^3 (4 k-1) (8 k-1) (8 k+1) (10 k-1)\cr
 &&\times (10 k+1) \left(12 k^5-168 k^4+384 k^3+478 k^2+154 k+15\right) \Big(1776 k^8+5396 k^7\cr
 &&+14776 k^6+20945 k^5+23464 k^4+13512 k^3+1880 k^2-778 k-121\Big) x^{6 k+4}\cr
 &&-25200 (1-8 k)^2 (1-4 k)^2 k (4 k+1)^3 (6 k-1) (6 k+1)^2 (8 k+1)^2 (10 k-1)\cr
&& \times(10 k+1) \left(8 k^3+12 k^2+24 k+11\right) x^{8 k+1}-151200 (1-2 k)^2 k (4 k-1) (4 k+1)^2
\cr
&&\times (6 k-1)(6 k+1)^2 (8 k-1) (8 k+1)^2 (10 k-1) (10 k+1) \left(2 k^2-12 k-5\right) \cr
&&\times \left(8 k^3+12 k^2+24 k+11\right) x^{8 k+2}-9450 (2 k-1)^3 (4 k+1) (6 k+1) (8 k-1)\cr
 &&\times (8 k+1)^2 (10 k-1) (10 k+1) \left(8 k^3+12 k^2+24 k+11\right) \Big(1056 k^7-14736 k^6\cr
 &&+40648 k^5+21908 k^4-6720 k^3-3072 k^2+206 k+85\Big) x^{8 k+3}-3150 (1-4 k)^2\cr
  &&\times (1-2 k)^4 (4 k+1) (6 k-1) (6 k+1) (10 k-1) (10 k+1) \left(8 k^3+12 k^2+24 k+11\right)\cr
   &&\times \left(288 k^7-6816 k^6+41288 k^5-42780 k^4-98488 k^3-49044 k^2-9238 k-585\right) x^{8 k+4}
\cr
&&+94500 k (4 k-1) (4 k+1)^3 (6 k-1) (6 k+1)^2 (8 k-1) (8 k+1) (10 k+1)^2 (1-2 k)^2 \cr
&&\times(1-10 k)^2 \left(2 k^2-12 k-5\right) x^{10 k+1}+141750 k (2 k-1)^3 (4 k+1)^2 (6 k+1) (8 k-1)\cr
&&\times (8 k+1) (10 k-1) (10 k+1)^2 \Big(672 k^7-9168 k^6+25648 k^5+13092 k^4-4768 k^3\cr
&&-2044 k^2+138 k+55\Big) x^{10 k+2}+4725 (4 k-1) (4 k+1) (6 k+1) (10 k-1) (10 k+1)^2\cr
&&\times(1-2 k)^4 \Big(46080 k^{10}-1036800 k^9+6347040 k^8-7926032 k^7-13623936 k^6\cr
&&-3194472 k^5+1403580 k^4+601032 k^3+30696 k^2-10528 k-1035\Big) x^{10 k+3}\cr
&&+4725 (2 k-1)^5 (4 k+1) (6 k-1) (8 k-1) (1-4 k)^2 \Big(17280 k^{11}-586800 k^{10}\cr
&&+5939232 k^9-19042592 k^8-2829296 k^7+47470368 k^6+51882672 k^5+24555420 k^4\cr
&&+6245400 k^3+885440 k^2+65762 k+1989\Big) x^{10 k+4}-16 \Big(10970726400 k^{21}\cr
&&+146040422400 k^{20}+641482899456 k^{19}+1856477659136 k^{18}+4318539487232 k^{17}\cr
&&+8304191960320 k^{16}+13318866483328 k^{15}+17499377526912 k^{14}\cr
&&+18755721158592 k^{13}+15109498274704 k^{12}+9239954852200 k^{11}\cr
&&+4215820540256 k^{10}+1343741413744 k^9+260198810504 k^8+18391021108 k^7\cr
&&-6289646720 k^6-2254086318 k^5-335839017 k^4-29710472 k^3\cr
&&-1578494 k^2-25270 k-1\Big) x^5
\Big]\,,
\eea
and also
\bea
\phi &=& \epsilon (1-x^k) -\frac{\epsilon^3 k x^{k-1}}{24 \left(8 k^2-2 k-1\right)}\Big[\left(2 k^3-9 k^2+2\right) x^{2 k+1}+3 k (4 k+1) x^{2 k}\cr
&&-\left(2 k^3+3 k^2+3 k+2\right) x\Big] -\frac{\epsilon^5 k^2 x^{k-2}}{1920 \left(-8 k^2+2 k+1\right)^2 \left(24 k^2-2 k-1\right)}\Big[\cr
&&-30 k (k+1) (4 k-1) (4 k+1) (6 k+1) \left(2 k^2+k+2\right) x^{2 k+1}-10 (k+1) (2 k-1)\cr
&&\times(4 k-1)(6 k+1) \left(k^2-4 k-2\right) \left(2 k^2+k+2\right) x^{2 k+2}+45 k (2 k-1) (4 k+1)\cr
 &&\times (6 k+1) \left(4 k^3-20 k^2-2 k+3\right) x^{4 k+1}+3 (4 k-1) (4 k+1) (1-2 k)^2\cr
  &&\times \left(3 k^4-34 k^3+60 k^2+64 k+12\right) x^{4 k+2}+(k+1) \Big(384 k^7+880 k^6+844 k^5\cr
  &&+1196 k^4+495 k^3+433 k^2-28 k-4\Big) x^2+15 k (4 k-1) (4 k+1)^2 (5 k-1)\cr
  &&\times (6 k+1) x^{4 k}\Big] -\frac{\epsilon^7 k^3 x^{k-3}}{322560 \left(8 k^2-2 k-1\right)^3 \left(24 k^2-10 k+1\right) \left(48 k^2+14 k+1\right)}\Big[\cr
&&
-525 k (k+1) (4 k-1) (4 k+1)^2 (5 k-1) (6 k-1) (6 k+1) (8 k+1)\cr
&&\times\left(2 k^2+k+2\right) x^{4 k+1}-1575 k (k+1) (2 k-1) (4 k+1) (6 k-1) (6 k+1) (8 k+1)
 \cr
&&\times\left(2 k^2+k+2\right) \left(4 k^3-20 k^2-2 k+3\right) x^{4 k+2}+315 k (2 k-1) (4 k+1)^2 (6 k-1) \cr
&&\times(6 k+1)(7 k-1) (8 k+1) \left(8 k^3-43 k^2-2 k+7\right) x^{6 k+1}-105 (1-2 k)^2 (k+1)\cr
 &&\times(4 k-1) (4 k+1) (6 k-1) (8 k+1) \left(2 k^2+k+2\right) \left(3 k^4-34 k^3+60 k^2+64 k+12\right)
  \cr
&&\times x^{4 k+3}+21 k (k+1) (4 k+1) (6 k-1) (8 k+1) \Big(2112 k^7+4480 k^6+5452 k^5\cr
&&+6368 k^4+3015 k^3+1969 k^2-244 k-52\Big) x^{2 k+2}+7 (k+1) (2 k-1) (6 k-1)\cr
 &&\times(8 k+1) \left(k^2-4 k-2\right) \Big(2112 k^7+4480 k^6+5452 k^5+6368 k^4+3015 k^3+1969 k^2\cr
 &&-244 k-52\Big) x^{2 k+3}+315 k (4 k+1) (8 k+1) (1-2 k)^2 \Big(528 k^7-6240 k^6+13902 k^5\cr
 &&+8549 k^4-2480 k^3-1251 k^2+80 k+37\Big) x^{6 k+2}+15 (2 k-1)^3 (4 k-1) (4 k+1)\cr
  &&\times (6 k-1) \left(72 k^7-1449 k^6+7162 k^5-4290 k^4-14972 k^3-8316 k^2-1712 k-120\right)\cr
   &&\times x^{6 k+3}+105 k (4 k-1) (4 k+1)^3 (6 k-1) (6 k+1) (7 k-2) (7 k-1) (8 k+1) x^{6 k}\cr
   &&-(k+1) \Big(313344 k^{12}+1416960 k^{11}+2031488 k^{10}+2659344 k^9+3369536 k^8\cr
   &&+2520904 k^7+2329376 k^6+866389 k^5+542126 k^4+2059 k^3+13622 k^2\cr
   &&-156 k+8\Big) x^3\Big]\cr
    &&-\frac{\epsilon^9 k^4 x^{k-4}/92897280}{ \left(-24 k^2+2 k+1\right)^2 \left(-8 k^2+2 k+1\right)^4 \left(3840 k^4-256 k^3-124 k^2+4 k+1\right)}\Big[\nn\\
&&2835 k (3 k-1) (4 k+1)^4 (6 k-1) (6 k+1)^2 (8 k-1) (8 k+1) (9 k-2) (9 k-1)\cr
 &&\times (10 k+1) (1-4 k)^2 x^{8 k}-60 k (k+1) (4 k-1) (4 k+1) (6 k+1) (8 k-1) (10 k+1)\cr
&&\times\Big(1622016 k^{12}+6469632 k^{11}+11055040 k^{10}+15928848 k^9+17757472 k^8\cr
&&+14360408 k^7+10484068 k^6+4008509 k^5+1687990 k^4-210673 k^3-6314 k^2\cr
&&+2676 k+328\Big) x^{2 k+3}-20 (k+1) (2 k-1) (4 k-1) (6 k+1) (8 k-1)\cr
 &&\times(10 k+1) \left(k^2-4 k-2\right) \Big(1622016 k^{12}+6469632 k^{11}+11055040 k^{10}\cr
 &&+15928848 k^9+17757472 k^8+14360408 k^7+10484068 k^6+4008509 k^5\cr
 &&+1687990 k^4-210673 k^3-6314 k^2+2676 k+328\Big) x^{2 k+4}+630 k (k+1) (4 k-1)\cr
  &&\times (4 k+1)^2 (5 k-1) (6 k-1) (6 k+1) (8 k-1) (8 k+1) (10 k+1) \Big(3072 k^7\cr
  &&+6320 k^6+8372 k^5+9148 k^4+4545 k^3+2639 k^2-404 k-92\Big) x^{4 k+2}\cr
&&+1890 k (k+1) (2 k-1) (4 k+1) (6 k-1) (6 k+1) (8 k-1) (8 k+1) (10 k+1)\cr
&&\times\left(4 k^3-20 k^2-2 k+3\right) \Big(3072 k^7+6320 k^6+8372 k^5+9148 k^4+4545 k^3\cr
&&+2639 k^2-404 k-92\Big) x^{4 k+3}+126 (k+1) (4 k-1) (4 k+1) (6 k-1) (8 k-1)\cr
&&\times(8 k+1) (10 k+1) (1-2 k)^2 \left(3 k^4-34 k^3+60 k^2+64 k+12\right) \Big(3072 k^7\cr
&&+6320 k^6+8372 k^5+9148 k^4+4545 k^3+2639 k^2-404 k-92\Big) x^{4 k+4}\cr
&&-8820 (1-4 k)^2 k (k+1) (4 k+1)^3 (6 k-1) (6 k+1)^2 (7 k-2) (7 k-1) (8 k-1)\cr
&&\times (8 k+1) (10 k+1) \left(2 k^2+k+2\right) x^{6 k+1}-26460 k (k+1) (2 k-1) (4 k-1)\cr
 &&\times (4 k+1)^2 (6 k-1) (6 k+1)^2 (7 k-1) (8 k-1) (8 k+1) (10 k+1) \left(2 k^2+k+2\right)\cr
  &&\times \left(8 k^3-43 k^2-2 k+7\right) x^{6 k+2}-26460 (1-2 k)^2 k (k+1) (4 k-1) (4 k+1)\cr
&&\times(6 k+1) (8 k-1) (8 k+1) (10 k+1) \left(2 k^2+k+2\right) \Big(528 k^7-6240 k^6+13902 k^5\cr
&&+8549 k^4-2480 k^3-1251 k^2+80 k+37\Big) x^{6 k+3}-1260 (1-4 k)^2 (k+1) (2 k-1)^3\cr
 &&\times(4 k+1) (6 k-1) (6 k+1) (8 k-1) (10 k+1) \left(2 k^2+k+2\right) \Big(72 k^7-1449 k^6\cr
 &&+7162 k^5-4290 k^4-14972 k^3-8316 k^2-1712 k-120\Big) x^{6 k+4}+1890 k (2 k-1)\cr
  &&\times(4 k-1) (4 k+1)^3 (6 k-1) (6 k+1)^2 (8 k-1) (8 k+1) (9 k-2) (9 k-1) (10 k+1)\cr
   &&\times\left(20 k^3-112 k^2-2 k+19\right) x^{8 k+1}+2835 k (4 k+1)^2 (6 k+1) (8 k-1) (8 k+1)\cr
&&\times(9 k-1) (10 k+1) (1-2 k)^2 \Big(6720 k^8-82368 k^7+215376 k^6+48524 k^5\cr
&&-70874 k^4-5716 k^3+6472 k^2+140 k-149\Big) x^{8 k+2}+2835 k (2 k-1)^3 (4 k-1)\cr
 &&\times (4 k+1) (6 k+1) (10 k+1) \Big(153600 k^{11}-3068928 k^{10}+16559680 k^9-17930416 k^8\cr
 &&-28325608 k^7-1887768 k^6+4959928 k^5+865666 k^4-224292 k^3-45844 k^2\cr
 &&+2762 k+595\Big) x^{8 k+3}+315 (4 k+1) (6 k-1) (6 k+1) (8 k-1) (1-2 k)^4 (1-4 k)^2\cr
  &&\times \Big(960 k^{10}-28904 k^9+255619 k^8-687148 k^7-248660 k^6+1424192 k^5+1520776 k^4\cr
  &&+645008 k^3+135520 k^2+13952 k+560\Big) x^{8 k+4}+(k+1) \Big(17553162240 k^{19}\cr
  &&+124438708224 k^{18}+271959490560 k^{17}+337797214208 k^{16}+528200379392 k^{15}\cr
  &&+626970144000 k^{14}+599490141888 k^{13}+573004064640 k^{12}+344069871216 k^{11}\cr
  &&+252342317824 k^{10}+75725482900 k^9+40504057176 k^8-2466846655 k^7\cr
  &&+594782927 k^6-490948197 k^5-88246935 k^4-1565208 k^3\cr
  &&-1021080 k^2+10864 k+16\Big) x^4\Big]\,.
\eea
The mass is given by (\ref{mass}). Electric and magnetic potentials are given by
\bea
&&\Phi_p + \Phi_q = 2\sqrt2 + \frac{\epsilon^2 (k-1) k}{4 \sqrt{2}} +
\frac{\epsilon^4 (k-1) k^2 \left(4 k^2+5 k-17\right)}{384 \sqrt{2} (4 k+1)}\cr
&&+\frac{\epsilon^6 (k-1) k^3 \left(96 k^5+256 k^4-482 k^3-627 k^2+1636 k+721\right)}{92160 \sqrt{2} (4 k+1)^2 (6 k+1)}\cr
&&+\frac{\epsilon^8 (k-1) k^4/41287680}{ \sqrt{2} (4 k+1)^3 (6 k+1) (8 k+1)}\Big[
3072 k^8+12416 k^7-28256 k^6-229080 k^5\cr
&&-128126 k^4+49307 k^3-218013 k^2-209383 k-58337\Big]\cr
&&+\frac{\epsilon^{10} (k-1) k^5/29727129600}{ \sqrt{2} (4 k+1)^4 (6 k+1)^2 (8 k+1) (10 k+1)}
\Big[737280 k^{12}+4098048 k^{11}+18086912 k^{10}\cr
&&+359104768 k^9+1348868928 k^8+2065662408 k^7+1359107628 k^6+120145038 k^5\cr
&&-96852447 k^4+189495288 k^3+193902858 k^2+74709050 k+7734241
\Big] + {\cal O}(\epsilon^{12})\,,\nn\\
&&\Phi_p-\Phi_q =\frac{\epsilon (k-1) k}{\sqrt{k (k+1)}} + \frac{\epsilon^3 (k-1) k^2 \left(4 k^2+5 k-5\right)}{48 \sqrt{k (k+1)} (4 k+1)}\cr
&&+\frac{\epsilon^5 (k-1) k^3 \left(96 k^5+256 k^4-482 k^3-867 k^2+76 k+121\right)}{7680 \sqrt{k (k+1)} (4 k+1)^2 (6 k+1)}\cr
&& + \frac{\epsilon^7 (k-1) k^4/2580480}{ \sqrt{k (k+1)} (4 k+1)^3 (6 k+1) (8 k+1)}
\Big[3072 k^8+12416 k^7+4000 k^6-24792 k^5\cr
&&+113794 k^4+224531 k^3+83799 k^2-6775 k-6845\Big]\cr
&&+\frac{\epsilon^9 (k-1) k^5/1486356480 }{\sqrt{k (k+1)} (4 k+1)^4 (6 k+1)^2 (8 k+1) (10 k+1)}
\Big[737280 k^{12}+4098048 k^{11}\cr
&&-26149888 k^{10}-195329792 k^9-391234752 k^8-426620472 k^7-528488532 k^6\cr
&&-645596082 k^5-449469567 k^4-153041592 k^3-16272102 k^2+4269290 k+698161\Big]\nn\\
&&+ {\cal O}(\epsilon^{11})\,.
\eea

\section{Numerical procedure}
\label{app:numeric}

In this appendix, we describe our numerical procedure for constructing
the extremal dyonic black holes in the EMD theory.  The equations of
motion for the extremal case are reduced to (\ref{extremaleom})
and (\ref{extremalcons}). In order to perform the numerical calculations,
we need to set up the boundary data. Since the equations become
singular on
the horizon itself, we choose to set the integration boundary data
slightly outside the horizon.  Assuming that the horizon is at $r_0$,
for given $p$ and $q$, we can obtain the near-horizon solution analytically
by means of power expansions in $(r-r_0)$.  By working to
sufficiently high order in powers of $(r-r_0)$,
this approximate solution allows us to set initial data for the
numerical integration of the system from near horizon to asymptotic infinity
and to achieve the desired degree of numerical accuracy.
 From the resulting numerical solution we can then
read off the asymptotic data such as, in particular, the mass of the extremal
dyonic black hole.

The charge parameters are given by
\be
q= \sqrt2 r_0 e^{\fft12 a \phi_0} \,, \qquad p = \sqrt2 r_0 e^{-\fft12 a \phi_0} \,.
\ee
In other words, for extremal dyonic black holes, the solution is
specified by the electric and magnetic charges, which are related to
the horizon radius $r_0$ and the value $\phi_0=\phi(r_0)$ of
the dilaton $\phi(r)$ on the horizon.  The entropy is given
by $S=\pi r_0^2$.  With the dilaton coupling constant $a$
parameterized by (\ref{asqandx}), we find that the leading order of
the near-horizon expansion is
\be
h=h_2 (r-r_0)^2 + \cdots\,,\qquad
\phi=\phi_0 + \phi_k (r-r_0)^k + \cdots\,.
\ee
In this paper, we shall focus on the cases where $k$ is integer, so
that the series expansion is analytic.
For low-lying examples of $k$, we find
\bea
&&\qquad a=\sqrt3\,,\quad k=2:\nn\\
\fft{h}{h_2} &=& \sum_{i=2} \fft{i-1}{r_0^{i-2}} (\delta r)^i  + \ft16 r_0 \phi_2^2 (\delta r)^5 + \cdots\,,\nn\\
\phi &=&\phi_0 + \phi_2 (\delta r)^2 - \fft{2\phi_2}{r_0} (\delta r)^3 - \fft{3\phi_2}{r_0^2} (\delta r)^4
-\big(\frac{4 \phi_2}{r_0^3}-\frac{r_0 \phi_2^3}{6}\big)(\delta r)^5 +
\cdots\,,\nn\\  \nn\\
&&\qquad a=\sqrt6\,,\quad k=3:\nn\\
\fft{h}{h_2} &=&\sum_{i=2} \fft{i-1}{r_0^{i-2}} (\delta r)^i + \ft3{20} r_0 \phi_3^2 (\delta r)^7 +
\cdots\,,\nn\\
\phi &=& \phi_0 + \phi_3 (\delta r)^3 - \fft{3\phi_3}{r_0} (\delta r)^4 +
\fft{6\phi_3}{r_0^2} (\delta r)^5 -\fft{10\phi_3}{r_0^3} (\delta r)^6 +
\fft{15\phi_3}{r_0^4} (\delta r)^7 +\cdots\,,\nn\\ \nn\\
&&\qquad a=\sqrt{10}\,,\quad k=4:\nn\\
\fft{h}{h_2} &=&\sum_{i=2} \fft{i-1}{r_0^{i-2}} (\delta r)^i + \ft17{r_0 \phi_4^2} (\delta r)^9 +
\cdots\,,\nn\\
\phi &=& \phi_0 + \phi_4 (\delta r)^4 - \fft{4\phi_4}{r_0} (\delta r)^6 + \fft{10\phi_4}{r_0^2} (\delta r)^6 -\fft{20\phi_4}{r_0^4} (\delta r)^6 + \fft{35\phi_4}{r_0^4} (\delta r)^8 +\cdots\,,
\eea
where $\delta r=r-r_0$.  There are two free parameters that
are to be determined, namely
$h_2$ and $\phi_k$.  The parameter $h_2$ is fixed by requiring
that $h(r\rightarrow \infty)=1$.  The parameter $\phi_k$ is fixed by
requiring that $\phi(r\rightarrow \infty)=0$. After obtaining the
numerical solutions, the mass can be read off from the formula
\be
M = \ft12 r^2 h'\Big|_{r\rightarrow \infty}\,.
\ee

\end{document}